\documentclass[10pt,twocolumn,twoside,submit]{JCNtran}

\usepackage[dvips]{epsfig}
\usepackage[latin1]{inputenc}
\usepackage[T1]{fontenc}

\def\BibTeX{{\rm B\kern-.05em{\sc i\kern-.025em b}\kern-.08em
    T\kern-.1667em\lower.7ex\hbox{E}\kern-.125emX}}

\usepackage{fancyhdr}
\usepackage{acronym}
\usepackage{graphicx}
\usepackage{amsfonts}
\usepackage{times}
\usepackage{latexsym}
\usepackage{amssymb}
\usepackage{amsmath}
\usepackage{cite}
\usepackage{verbatim}
\usepackage{subfigure}
\usepackage{multirow}
\usepackage{lastpage}
\usepackage{calc}
\usepackage{eso-pic}
\usepackage{tabularx}
\usepackage{ifthen}

\DeclareGraphicsExtensions{.pdf}

\newcommand{\tr}{\mathop{\mathrm{tr}}\nolimits}

\newcommand{\figref}[1]{{Fig.}~\ref{#1}}
\newcommand{\secref}[1]{{Section}~\ref{#1}}

\newcommand{\bydef}{\triangleq}

\newtheorem{Theorem}{Theorem}

\newtheorem{Lemma}{Lemma}

\def\cB{{\mathcal{B}}}
\def\cC{{\mathcal{C}}}

\def\cN{{\mathcal{N}}}

\def\cU{{\mathcal{U}}}

\def\bpsi{{\boldsymbol{\psi}}}

\def\bPsi{{\boldsymbol{\Psi}}}
\def\bPhi{{\boldsymbol{\Phi}}}

\def\bTheta{{\boldsymbol{\Theta}}}

\def\bydef{:=}

\def\bee{{\mathbf{e}}}
\def\bff{{\mathbf{f}}}
\def\bg{{\mathbf{g}}}
\def\bh{{\mathbf{h}}}

\def\bv{{\mathbf{v}}}
\def\bw{{\mathbf{w}}}
\def\bx{{\mathbf{x}}}
\def\by{{\mathbf{y}}}
\def\bz{{\mathbf{z}}}
\def\b0{{\mathbf{0}}}

\def\bA{{\mathbf{A}}}

\def\bF{{\mathbf{F}}}

\def\bH{{\mathbf{H}}}
\def\bI{{\mathbf{I}}}

\def\bQ{{\mathbf{Q}}}
\def\bR{{\mathbf{R}}}

\def\bT{{\mathbf{T}}}

\def\bV{{\mathbf{V}}}
\def\bW{{\mathbf{W}}}

\def\bY{{\mathbf{Y}}}
\def\bZ{{\mathbf{Z}}}

\def\bbC{{\mathbb{C}}}

\def\bbE{{\mathbb{E}}}

\def\rD{{\mathrm{D}}}

\def\ra{{\mathrm{a}}}

\def\rc{{\mathrm{c}}}

\def\rf{{\mathrm{f}}}

\def\rp{{\mathrm{p}}}

\def\rr{{\mathrm{r}}}

\def\rt{{\mathrm{t}}}

\def\rx{{\mathrm{x}}}
\def\ry{{\mathrm{y}}}

\def\bydef{:=}

\def\sf0{{\mathsf{0}}}

\def\ras{{\mathrm{as}}}

\def\var{{\mathrm{var}}}

\newcommand{\hatvh}{\hat{\bh}} 
\newcommand{\barvh}{\bar{\bh}}

\newcommand{\brevevh}{\breve{\bh}}

\newcommand{\barvR}{\bar{\bR}}

\def\nn{\nonumber}

\pubyear{08}{2013}

\begin{document}
\bibliographystyle{jcn}

\title{Effects of Channel Aging in Massive MIMO Systems}
\author{Kien T. Truong and Robert W. Heath Jr.
\thanks{Manuscript received February 28, 2013; approved for publication by Dr. Giuseppe Caire, Associate Editor, June 17, 2013.}
\thanks{This research was supported by Huawei Technologies.}
\thanks{Kien T. Truong is with MIMO Wireless Inc, Austin, TX 78704, email: kientruong@utexas.edu}
\thanks{Robert W. Heath Jr. is with The University of Texas at Austin, Austin, TX 78712, email: rheath@utexas.edu. He is also President and CEO of MIMO Wireless Inc. The terms of this arrangement have been reviewed and approved by the University of Texas at Austin in accordance with its policy on objectivity in research.}}\markboth{JOURNAL OF
COMMUNICATIONS AND NETWORKS, VOL. ..., NO. ..., AUGUST
2003}{Truong and Heath: Effects of Channel Aging...} \maketitle

\begin{abstract}
MIMO communication may provide high spectral efficiency through the deployment of a very large number of antenna elements at the base stations. The gains from massive MIMO communication come from the use of multi-user MIMO on the uplink and downlink, but with a large excess of antennas at the base station compared to the number of served users. Initial work on massive MIMO did not fully address several  practical issues associated with its deployment. This paper considers the impact of channel aging on the performance of massive MIMO systems. The effects of channel variation are characterized as a function of different system parameters assuming a simple model for the channel time variations at the transmitter. Channel prediction is proposed to overcome channel aging effects. The analytical results on aging show how capacity is lost due to time variation in the channel. Numerical results in a multicell network show that massive MIMO works even with some channel variation and that channel prediction could partially overcome channel aging effects.
\end{abstract}

\begin{keywords}
Massive MIMO, large-scale antenna systems, channel aging, outdated CSI, channel prediction.
\end{keywords}

\section{\uppercase{Introduction}}
\label{sec:introduction}

Massive multiple-input multiple-output (MIMO) is a new breakthrough communication technique. The key ideas of massive MIMO are to deploy a very large number of antennas at each base station and to use multi-user MIMO (MU-MIMO) transmission to serve a much smaller number of users~\cite{Marzetta2010:TWC, RusekEtAl2012:SPM} 
In a typical envisioned deployment scenario, each base station has hundreds of antennas to simultaneously serve tens of single-antenna users. The large excess of antennas at the base station makes it possible to design low-complexity linear signal processing strategies that are well matched to the propagation channel to maximize system capacity. Although the theory of massive MIMO is now established (see~\cite{RusekEtAl2012:SPM} and references therein) and preliminary system level simulations are promising~\cite{Samsung2012:Slides}, further investigation under practical settings is needed to understand the real potential of this technique. 

Prior work on massive MIMO communication considers the impact of channel estimation error due to noise or pilot contamination.  An important observation in prior work is that pilot contamination puts deterministic limits on the signal-to-interference-plus-noise ratio (SINR) and hence the achievable rates~\cite{AppaiahEtAl2010:ICC,AshikhminMarzetta2012:ISIT,FernandesEtAl2012:ICC,GopalakrishnanJindal2011:SPAWC,JoseEtAl2011:TWC,NgoEtAl2011:ICASSP}. In addition to estimation errors, another reason for channel state information (CSI) inaccuracy is channel aging. Due to time variation of the propagation channel and delays in the computation, the channel varies between when it is learned at the base station and when it is used for beamforming or detection. The impact of channel aging has not yet been fully characterized in prior work on massive MIMO. Although channel aging has been studied in other MIMO cellular configurations,  like in multicell transmission~\cite{ThieleEtAl2011:Asilomar}, such results are not directly applicable to massive MIMO systems. 

In this paper, we incorporate the practical impairment known as channel aging into massive MIMO systems on both the uplink and the downlink. For performance analysis, we adopt the approach using deterministic equivalents developed in~\cite{HoydisEtAl2011:Allerton, HoydisEtAl2013:JSAC} for cellular networks where the number of antennas at each base station is much larger than the number of active users per cell. Although the existing framework in~\cite{HoydisEtAl2011:Allerton, HoydisEtAl2013:JSAC} allows for taking into account certain practical effects like antenna correlation, their main focus is to develop an analytical framework based on random matrix theory.  By introducing time variation into the framework developed in~\cite{HoydisEtAl2011:Allerton, HoydisEtAl2013:JSAC}, our results are a natural, but not straightforward, generalization of those in~\cite{HoydisEtAl2011:Allerton, HoydisEtAl2013:JSAC}. Specifically, we provide asymptotic analysis on the impact of channel aging on both the uplink and the downlink achievable rates when the maximal ratio combining (MRC) receiver or the matched filtering (MF) precoder is used. The analysis allows for the characterization of the performance loss due to channel aging. Our analysis shows that channel aging mainly affects the desired signal power to a user and the inter-cell interference due to pilot contamination (corresponding to users in other cells that share the same pilot as the user). We also report on one approach for mitigating channel aging effects based on the finite impulse response (FIR) Wiener predictor. The idea is to use current and past observations to predict future channel realizations and thus reduce the impact of aging. We incorporate prediction into the deterministic equivalent analysis under some assumptions. Our work provides a foundation for incorporating better predictors into massive MIMO in the future.

We numerically investigate channel aging effects and channel prediction benefits in a multi-cell massive MIMO network with realistic parameters. Our simulation results show how channel aging degrades the performance of massive MIMO systems on both uplink and downlink. Notably, our results show that massive MIMO still works even when there is some time variation in the channel. For example, the achievable rate in the aged CSI case is still about half of that in the current CSI case if the normalized Doppler shifts are as large as $0.2$ on the uplink and on the downlink. Our results also show that by exploiting temporal correlation in the channel the proposed linear FIR channel predictor could partially overcome the effects of aging, though further work is needed to fully investigate the potential of prediction. 

The remainder of this paper is organized as follows. \secref{sec:systemModel} describes the system model. \secref{sec:channelAging} generalizes the framework in \cite{HoydisEtAl2011:Allerton, HoydisEtAl2013:JSAC} to incorporate channel aging and shows how to overcome channel aging based on linear FIR prediction. \secref{sec:analysis} provides an analysis of the achievable rates on the uplink and on the downlink in the presence of channel aging and/or channel prediction. \secref{sec:numericalResults} numerically investigates the effects of channel aging and the benefits of channel prediction in a multicell network. \secref{sec:conclusion} concludes the paper and provides suggestions for future work.

\vspace{10pt}
\section{\uppercase{System Model}}
\label{sec:systemModel}
Consider a cellular network with $C$ cells. Each cell has a base station and $U$ randomly distributed active users. Let $\cC \bydef \{ 1, 2, \cdots, C\}$ be the set of indices of the cells. Let $\cU_{c}\bydef \{1, 2, \cdots, U\}$ be the set of indices of active users in cell $c \in \cC$ and $\cU \bydef \cU_{1}\cup \cU_{2}\cup \cdots \cup \cU_{C}$ be the set of indices of all active users in the network. Each base station is equipped with $N_{\rt}$ antennas and each active user is equipped with a single antenna. A distinguishing feature of massive MIMO systems is that the number of antennas at each base station is much larger than the number of served users, i.e. $N_{\rt} \gg U \gg 1$. The network operates in a time division duplex (TDD) protocol, i.e. each node uses a single frequency for both transmission and reception of signals. Since each node cannot transmit and receive on the same frequency at the same time, the transmission and reception at each node are spaced apart by multiplexing signals on a time basis. All base stations and active users are perfectly synchronized in time and frequency. As in 3GPP LTE/LTE-Advanced standards~\cite{LTEARequirements}, a frequency reuse of one is assumed. Extensions to other frequency reuse factors are straightforward~\cite{Marzetta2010:TWC}.

\begin{figure}[h]
\centerline{
\includegraphics[width=0.9\columnwidth]{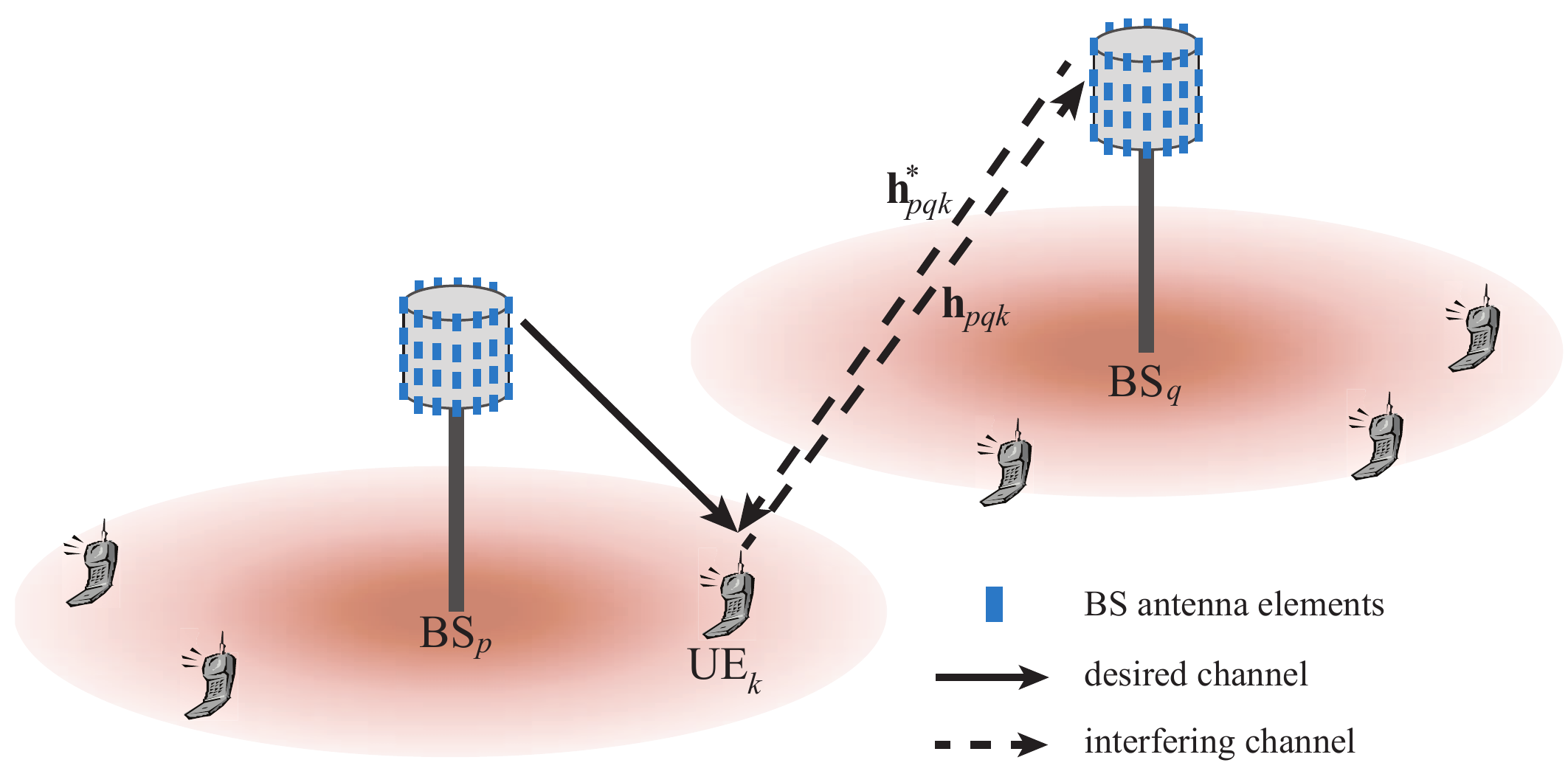}
}
\caption{The massive MIMO system under consideration. There are $C$ cells, each has one base station (BS) and $U$ single-antenna users. Each base station has $N_{\rt}$ antennas. $\bh_{bcu}[n]\in \bbC^{N_{\rt}}$ is the column channel vector from user $u$ in cell $c$ to base station $b$ at time $n$. Due to channel reciprocity, we assume that $\bh^{*}_{bcu}[n]$ is the channel vector from base station $b$ to user $u$ in cell $c$ at time $n$.}
\label{fig:systemModel}
\end{figure}

We assume that the channels are frequency flat, i.e. a single frequency band or subcarrier; the extension to OFDM-based frequency selective channels with multiple subcarriers follows in a similar manner. We consider a quasi-static block fading channel model where the channel bandwidth is much smaller than the coherence bandwidth and the channel coefficients do not change within one symbol, but vary from symbol to symbol. Let $\bh_{bcu}[n] \in \bbC^{N_{\rt} \times 1}$ be the channel vector from user $u$ in cell $c$ to base station $b$ at the $n$-th symbol. Define $\bH_{bc}[n]\bydef [\bh_{bc1}[n],~\bh_{bc2}[n], \cdots, \bh_{bcU}[n] ] \in \bbC^{N_{\rt} \times U}$ as the combined channel matrix from all users in cell $c$ to base station $b$. For analysis, we assume that $\bh_{bcu}[n]$ is modeled as~\cite{HoydisEtAl2011:Allerton,HoydisEtAl2013:JSAC}
\begin{align}\label{eq:channelModel}
\bh_{bcu}[n] 
\bydef& \bR^{1/2}_{bcu}\bg_{bcu}[n],
\end{align}
where $\bg_{bcu}[n] \in \bbC^{N_{\rt}}$ is a fast fading channel vector and $\bR_{bcu} \in \bbC^{N_{\rt} \times N_{\rt}}$ is a deterministic Hermitian-symmetric positive definite matrix. We assume that $\bg_{bcu}[n]$ is uncorrelated wide-sense stationary complex Gaussian random processes with zero mean and unit variance, i.e. $\bg_{bcu}[n] \sim \cC\cN(\b0,\bI_{N_{\rt}})$. The deterministic matrix is independent of symbol index $n$ and is determined as
\begin{align}
\bR_{bcu} = & \bbE[\bh_{bcu}[n] \bh^{*}_{bcu}[n]],
\end{align}
for all $n$. $\bR_{bcu}$ may include many effects like pathloss, shadowing, building penetration losses, spatial correlation, and antenna patterns.  \figref{fig:systemModel} illustrates the system model under consideration. 


On the uplink (or reverse link), the users simultaneously send data to their serving base stations. Let $x_{\rr,cu}[n]$ be the transmitted symbol sent by user $u$ in cell $c \in \cC$ on the uplink at time $n$, where $\bbE[|x_{\rr,cu}[n]|^2] = 1$. The subscript $\rr$ is used to denote \underline{r}everse link. The transmitted symbols sent by the users are mutually independent. Define $\bx_{\rr,c}[n] \bydef [x_{\rr,c1}[n],~x_{\rr,c2}[n],\cdots, x_{\rr,cU}[n]]^{T} \in \bbC^{U}$ as the transmitted symbol vector by the $U$ users in cell $c$. The users use the same average transmit power of $p_{\rr}$ during the uplink data transmission stage. Let $\bz_{\rr,b}[n]\in \bbC^{N_{\rt}}$ be spatially white additive Gaussian noise at base station $b$, where $\bz_{r,b}[n] \sim \cC\cN(\b0,\sigma_{b}^{2}\bI_{N_{\rt}})$. Base station $b$ observes
\begin{align}
\by_{\rr,b}[n]
= & \sqrt{p_{\rr}}\sum_{c=1}^{C} \bH_{bc}[n]\bx_{\rr,c}[n] + \bz_{\rr,b}[n].
\end{align}
Base station $b$ applies a linear detector $\bW_{b}[n] \in \bbC^{N_{\rt} \times U}$ to $\by_{\rr,b}[n]$ to detect $\bx_{\rr,c}[n]$, where the $u$-th column of $\bW_{b}[n]$ is denoted as $\bw_{bu}[n]$. Denote  $\tilde{z}_{\rr,bu} = \bw_{\rr,bu}^{*}[n]\bz_{\rr,b}[n]$ as spatially filtered Gaussian noise. The post-processing received signal for detecting $x_{\rr,cu}[n]$ is 
\begin{align}
\tilde{y}_{\rr, bu}[n] 
= & \underbrace{\bw_{bu}^{*}[n]\bh_{bbu}[n]x_{\rr,bu}[n]}_{\mbox{desired~signal}}+ \underbrace{\frac{1}{\sqrt{p_{\rr}}}\tilde{z}_{\rr,bu}[n]}_{\mbox{noise}}\nn\\
& + \underbrace{\sum_{(c,k) \neq (b,u)} \bw_{bu}^{*}[n]\bh_{bck}[n]x_{\rr,ck}[n]}_{\mbox{interference}}.\label{eq:postprocessedULSignal}
\end{align}

On the downlink (or forward link), each base station uses MU-MIMO transmission strategies to broadcast data to its associated users. Since the base stations simultaneously send data, downlink transmission forms an interfering broadcast channel. The subscript $\rf$ is used to denote \underline{f}orward link. Denote $\bx_{\rf,b}[n] \bydef [x_{\rf,b1}[n],~x_{\rf,b2}[n],\cdots,~x_{\rf,bU}[n]]^{T} \in \bbC^{U}$ as the data symbols that base station $b$ sends to its serving $U$ users, where $\bbE\left[\bx_{\rf,b}[n]\right] = \b0$ and $\bbE\left[\bx_{\rf,b}[n]\bx^{*}_{\rf,b}[n]\right] = \bI_{N_{\rt}}$. Base station $b$ uses a linear precoding matrix $\bF_{b}[n] \in \bbC^{N_{\rt} \times U}$ to map $\bx_{\rf,b}[n]$ to its transmit antennas. The signal vector transmitted by this base station is $\sqrt{\lambda_{b}}\bF_{b}[n]\bx_{\rf,b}[n]$, where $\lambda_{b}$ is the normalization factor to satisfy the average transmit power constraint
\begin{align}
\lambda_{b} \bydef & \frac{1}{\bbE [\tr\bF_{b}[n]\bF^{*}_{b}[n]]}.\label{eq:lambda}
\end{align}
The base stations use the same average transmit power of $p_{\rf}$ during the downlink data transmission stage.  Let $z_{\rf,bu}[n] \sim \cC\cN(0,\sigma_{bu}^{2})$ be complex Gaussian noise at user $u$ in cell $b$. Define the combined noise vector at the users in cell $b$ as $\bz_{\rf,b}[n] \bydef [z_{\rf,b1}[n],~z_{\rf,b2}[n],~\cdots,~z_{\rf,bU}[n]]^{T} \in \bbC^{U}$. Let $\bff_{bu}[n]$ be the $u$-th column of $\bF_{b}[n]$. User $u$ in cell $b$ observes
\begin{align}
y_{\rf,bu}[n]
= &\underbrace{\sqrt{p_{\rf}}\sqrt{\lambda_{b}}\bh^{*}_{bbu}[n]\bff_{bu}[n]x_{\rf,bu}[n]}_{\mbox{desired~signal}}+ \underbrace{z_{\rf,bu}[n]}_{\mbox{noise}}\nn\\
&+ \underbrace{\sum_{(c,k) \neq (b,u)} \sqrt{p_{\rf}}\sqrt{\lambda_{c}}\bh^{*}_{cbu}[n]\bff_{ck}[n]x_{\rf,ck}[n]}_{\mbox{interference}}.\label{eq:DLreceivedSignal}
\end{align}

The base stations estimate the channels based on pilots, or training sequences, sent by the users. Let $\tau$ be the length of training period. The subscript $\rp$ is used to denote the \underline{p}ilot transmission stage, or the training stage. Suppose that all cells share the same set of $U$ pair-wisely orthogonal pilot signals $\bPsi \bydef [\bpsi_{1}; \cdots; \bpsi_{U}] \in \bbC^{U \times \tau}$, where $\bpsi_{u} \in \bbC^{1 \times \tau}$ for $u  = 1, \cdots, U$. The training sequences are normalized so that $\bPsi\bPsi^{*} = \bI_U$. The users use the same average transmit power of $p_{\rp}$ the training stage. The received training signal at base station $b$ is
\begin{align}\label{eq:Ypt}
\bY_{\rp,b}[n]
= & \sqrt{p_{\rp}\tau}\bigg(\sum_{c=1}^{C} \bH_{bc}[n]\bigg)\bPsi + \bZ_{\rp,b}[n],
\end{align}
where $\bZ_{\rp,b}[n] \in \bbC^{N_{\rt} \times \tau}$ is spatially white additive Gaussian noise matrix at base station $b$ during the training stage. Base station $b$ correlates $\bY_{\rp,b}[n]$ with $\bPsi$ to obtain
\begin{align}\label{eq:Ypt2}
\tilde{\bY}_{\rp,b}[n]
= & \frac{1}{\sqrt{p_{\rp}\tau}}\bY_{\rp,b}[n]\bPsi^{*}.
\end{align}
This gives the following noisy observation of the channel vector from user $u \in \cU_{b}$ to base station $b$
\begin{align}
\tilde{\by}_{\rp,bu}[n]
= & \underbrace{\bh_{bbu}[n]}_{\mbox{desired}} + \underbrace{\sum_{c \neq b} \bh_{bcu}[n]}_{\mbox{interference}} + \underbrace{\frac{1}{\sqrt{p_{\rp}\tau}} \underbrace{\bZ_{\rp,b}[n]\bpsi^{*}_{u}}_{\tilde{\bz}_{\rp,b}[n]}}_{\mbox{noise}},\label{eq:Ypt3}
\end{align}
where $\tilde{\bz}_{\rp,b}[n] \sim \cC\cN(\b0,\sigma_{b}^{2}\bI_{N_{\rt}})$ is the post-processed noise at base station $b$. The interference channels during the training stage are those from the users in the other cells using the same pilot. The effect of these interference channels on channel estimation error and then on system performance is called pilot contamination~\cite{Marzetta2010:TWC}. Similarly, that of noise is called noise contamination. Note that this model for training and channel estimation in the presence of pilot contamination and noise contamination are proposed and used widely in prior work~\cite{NgoEtAl2011:ICASSP,HoydisEtAl2011:Allerton,HoydisEtAl2013:JSAC}.

Base station $b$ applies minimum mean square error (MMSE) estimation to the right-hand side of \eqref{eq:Ypt3} to estimate $\bh_{bbu}[n]$ for $u \in \cU_{b}$. Define $\barvR_{bu} \bydef \sum_{c=1}^{C}\bR_{bcu}$ as the sum of the covariance matrices from the base stations to user $u$ in cell $b$. The MMSE estimate of $\bh_{bbu}[n]$ is~\cite{Verdu1998:BOOK}
\begin{align}
\hat{\bh}_{bbu}[n]
= & \bR_{bbu}\bQ_{bu}\tilde{\by}_{\rp,bu}[n],\label{eq:MMSEchannelEstimate}
\end{align}
where
\begin{align}
\bQ_{bu}
= & \left(\frac{\sigma_{b}^{2}}{p_{\rp}\tau}\bI_{N_{\rt}} + \barvR_{bu}\right)^{-1}.
\end{align}
Since $\bR_{bcu}$ is independent of index $n$ for all $b \in \cB, c \in \cC$, and $u \in \cU_{c}$, then $\bQ_{bu}$ is independent of index $n$. To compute the channel estimates, base station $b$ needs to know the deterministic correlation matrices $\bR_{bbu}$ and $\barvR_{bu}$ for $u \in \cU_{b}$. The sum correlation can be estimated from the received signal using standard covariance estimation techniques; it is assumed to be known perfectly in the analysis along with $\bR_{bbu}$.  The distribution of $\hat{\bh}_{bbu}[n]$ is $\hat{\bh}_{bbu}[n] \sim \cC\cN(\b0,\bPhi_{bbu})$, where~\cite{HoydisEtAl2011:Allerton}
\begin{align}
\bPhi_{bcu}
= & \bR_{bbu}\bQ_{bu}\bR_{bcu}, \forall~b \in \cB, c\in\cC, u\in \cU_{c}.\label{eq:estimateCov}
\end{align}
By setting $c=b$ in \eqref{eq:estimateCov}, we obtain $\bPhi_{bbu}$. Note that $\bPhi_{bcu}$ is independent of index  $n$ for all $b \in \cB, c \in \cC$, and $u \in \cU_{c}$. Due to the orthogonality property of the MMSE estimation, the observed channel can be decomposed as
\begin{align}
\bh_{bbu}[n] = \hat{\bh}_{bbu}[n] + \tilde{\bh}_{bbu}[n],\label{eq:MMSEorthogonality}
\end{align}
where $\tilde{\bh}_{bbu}[n] \sim \cC\cN(\b0, \bR_{bbu} - \bPhi_{bbu})$ is the channel estimation error and is uncorrelated with $\hat{\bh}_{bbu}[n]$. Because $\tilde{\bh}_{bbu}[n]$ and $\hat{\bh}_{bbu}[n]$ are jointly Gaussian, they are statistically independent.

\vspace{10pt}
\section{\uppercase{Incorporating Channel Aging Effects}}
\label{sec:channelAging}
We present a method for incorporating channel aging effects into the existing framework in \secref{subsec:channelAgingModel}. We then derive the optimal linear FIR Weiner channel predictor to overcome the aging effects in \secref{subsec:channelPrediction}.

\subsection{Channel Aging}
\label{subsec:channelAgingModel}

In principle, the channel changes over time due to the movements of antennas and those of objects (or people) in the propagation medium. To analyze the impact of channel aging, we need a time-varying model for the channel. For simplification and tractability, we assume that the channel temporal statistics are the same for all antenna pairs. Further, we assume that every user moves with the same velocity, so that the time variation is not a function of the user index. While this is not practical, from an analysis perspective, the performance will be dominated by the user with the most varying channel. Consequently, we assume every user has the same (worst-case) variation.  Similar assumptions for channel temporal correlation are made in prior work on MIMO wireless channels~\cite{KermoalEtAl2002:JSAC,WeichselbergerEtAl2006:TWC,VuPaulraj2007:JSAC}.

Let $h[n]$ be the univariate random process modeling the fading channel coefficient from a base station antenna to a user antenna. We model the random process $h[n]$ as a complex Gaussian process with zero mean (we do not consider a line-of-sight component). Under this assumption, the time variation of the channel is completely characterized by the second order statistics of the channel, in particular the autocorrelation function of the channel, which is generally a function of propagation geometry, velocity of the user, and antenna characteristics~\cite{BaddourBeaulieu2005:TWC}. A commonly-used autocorrelation function is the Clarke-Gans model, which is often called the Jakes model and assumes that the propagation path consists of a two-dimensional isotropic scatter with a vertical monopole antenna at the receiver~\cite{Jakes1974:BOOK}. In this model, the normalized (unit variance) discrete-time autocorrelation of fading channel coefficients is~\cite{Jakes1974:BOOK}
\begin{align}
r_{h}[k] 
=& J_{0}(2 \pi f_{D}T_{s}|k|),\label{eq:scalarACF}
\end{align}
where $J_{0}(\cdot)$ is the zeroth-order Bessel function of the first kind, $T_{s}$ is the channel sampling duration, $f_{D}$ is the maximum Doppler shift, and $|k|$ is the delay in terms of the number of symbols. The maximum Doppler shift $f_{D}$ is given by
\begin{align}
f_{D}
=& \frac{vf_{c}}{c},
\end{align}
where $v$ is the velocity of the user in meters per second (mps), $c = 3 \times 10^{8}$\,mps is the speed of light, and $f_{c}$ is the carrier frequency. As the delay $|k|$ increases or the user moves faster, the autocorrelation $r_{h}[k]$ decreases in magnitude to zero though not monotonically since there are some ripples. Note that other models for the autocorrelation function can be used; the choice of \eqref{eq:scalarACF} primarily impacts the simulations.

To generate realizations of the channel model, we adopt the approach of using an autoregressive model of order $L$, denoted as AR$(L)$, for approximating the temporally correlated fading channel coefficient process $h[n]$~\cite{BaddourBeaulieu2005:TWC}. Specifically, we assume that 
\begin{align}
h[n]
=& \sum_{\ell =1}^{L} a_{\ell}h[n-\ell] + w[n],
\end{align}
where $\{a_{\ell}\}_{\ell=1}^{L}$ are the AR coefficients and $w[n]$ is temporally uncorrelated complex white Gaussian noise process with zero mean and variance $\sigma^{2}_{w,(L)}$. Given the desired autocorrelation functions $r_{h}[k]$ in \eqref{eq:scalarACF} for $k\geq 0$, we can use the Levinson-Durbin recursion to determine $\{a_{\ell}\}_{\ell=1}^{L}$ and $\sigma^{2}_{w,(L)}$. More details on how to simulate temporally correlated fading channels are referred to~\cite{BaddourBeaulieu2005:TWC}. Note that increasing the AR model order $L$ improves the accuracy of channel modeling but also increases the complexity of the associated analysis.

For analysis and to design simplified predictors, we use an AR$(1)$ approximate model for the fading channel coefficients. This allows us to incorporate channel aging into analysis that already includes channel estimation error. It is reasonable to design predictors based on the AR$(1)$ model because it only requires estimating the parameters of the AR$(1)$ model; designing more elaborate predictors is a topic of future work. We denote $\alpha = J_{0}(2\pi f_{D}T_{s})$ as a temporal correlation parameter that corresponds to $r_{h}[1]$ in \eqref{eq:scalarACF}. We assume that $\alpha$ is known perfectly at the base stations. Let $\bh_{bcu}[n]$ be the channel vector between a user and a base station at time $n$. Under the AR$(1)$ model, for any $b, c \in \cC$ and $u \in \cU_{c}$,
\begin{align}
\bh_{bcu}[n]  =& \alpha \bh_{bcu}[n-1] + \bee_{bcu}[n],\label{eq:GaussMarkoModel}
\end{align}
where $\bh_{bcu}[n-1]$ is the channel in the previous symbol duration and $\bee_{bcu}[n] \in \bbC^{N_{\rt}}$ is an uncorrelated channel error due to channel aging. We assume that $\bee_{bcu}[n]$ is uncorrelated with $\bh_{bcu}[n-1]$ and is modeled as a stationary Gaussian random process with i.i.d.~entries and distribution $\cC\cN(\b0,(1-\alpha^2)\bR_{bcu})$~\cite{VuPaulraj2007:JSAC}. Note that the channel model in \eqref{eq:GaussMarkoModel} is also known as the stationary ergodic Gauss-Markov block fading channel model and has been used in prior work on multiuser MIMO~\cite{ZhangEtAl2009:JASP,CaireEtAl2010:TIT,AdhikaryEtAl2011:Allerton}. It follows from~\eqref{eq:GaussMarkoModel} that
\begin{align}
\bbE[\bh_{bcu}[n-q]\bh^{*}_{bcu}[n-k]] =& \alpha^{|k-q|}\bR_{bcu}.\label{eq:autoCorrelation2}
\end{align}

Now we establish a model for the combined effects of channel estimation errors and aging. We denote $n$ as the index of the channel sample where the channel is estimated. This means that based on the channel estimate $\hatvh_{bbu}[n]$ for all users $u \in \cU_{b}$, base station $b\in \cB$ designs the precoder $\bF_{b}[n+D]$ or the decoder $\bW_{b}[n+D]$, which is actually used at time $(n+D)$. For illustration, we assume that the CSI at the base stations is outdated by a channel sample duration, i.e. $D=1$. We refer to this as a one frame delay; extensions to larger delays, i.e. $D > 1$, are straightforward. It follows from \eqref{eq:MMSEorthogonality} and  \eqref{eq:GaussMarkoModel} that the true channel at time $(n+1)$ can be decomposed as 
\begin{align}
\bh_{bbu}[n+1] 
= & \alpha \bh_{bbu}[n] + \bee_{bbu}[n+1]\label{eq:GaussMarkov1}\\
= & \alpha\hat{\bh}_{bbu}[n] + \underbrace{\alpha \tilde{\bh}_{bbu}[n] +  \bee_{bbu}[n+1]}_{\tilde{\bee}_{bbu}[n+1]},\label{eq:GaussMarkov2}
\end{align}
where $\tilde{\bee}_{bbu}[n+1]\sim \cC\cN(\b0, \bR_{bbu} - \alpha^{2}\bPhi_{bbu})$ is mutually independent of $\hatvh_{bbu}[n]$. Note that while pilot contamination affects the estimation error $\tilde{\bh}_{bbu}[n] $, mobility and processing delay affect the aging error $\bee_{bbu}[n+1]$. 

\subsection{Channel Prediction}
\label{subsec:channelPrediction}
Channel prediction is one natural approach to overcome the channel aging effects. In this section, we focus on predicting $\bh_{bbu}[n+1]$ based on the current and previous received training signals. Effectively, we have the problem of predicting an autoregressive multivariate random process in the presence of noise. For simplicity, we assume that the interference from other base stations during training periods can be treated as uncorrelated additive Gaussian noise with zero mean and \emph{constant} variance. In practice, these interference channels change over time as the user moves. 
Let $\{\bV_{bbu,q}\}_{q=0}^{p}$, where $\bV_{bbu,q} \in \bbC^{N_{\rt} \times N_{\rt}}$, be the optimal $p$-th order Wiener linear predictor that minimizes the mean square error (MSE) in the prediction of $\bh_{bbu}[n+1]$ based on $\tilde{\by}_{\rp,bu}[n], \tilde{\by}_{\rp,bu}[n-1], \cdots, \tilde{\by}_{\rp,bu}[n-p]$. For notational convenience, we define $\bV_{bbu} \bydef [\bV_{bbu,0}~\bV_{bbu,1}~\cdots~\bV_{bbu,p}] \in \bbC^{N_{\rt}\times N_{\rt}(p+1) }$ and $\bar{\by}_{\rp, bu}[n] \bydef [\tilde{\by}^{*}_{\rp,bu}[n]~\tilde{\by}^{*}_{\rp,bu}[n-1]~\cdots~\tilde{\by}^{*}_{\rp,bu}[n-p]]^{*} \in \bbC^{N_{\rt}(p+1) \times 1}$. For $b, c \in \cC$ and $u \in \cU_{b}$, define 
\begin{align}
\delta(p,\alpha) \bydef&[1~\alpha~\cdots~\alpha^{p}]\\
\Delta(p,\alpha) \bydef& \begin{pmatrix}
1 		&	\alpha 	&	\cdots 	&	\alpha^{p}\\
\alpha 	&	1		&	\cdots	& 	\alpha^{p-1}\\
\vdots	&	\vdots	&	\ddots	&	\vdots\\
\alpha^{p} &	\alpha^{p-1} &	\cdots	&	1
\end{pmatrix}\\
\bT_{bu}(p,\alpha) 
\bydef& \left[\Delta(p,\alpha)\otimes \barvR_{bu} + \frac{\sigma_{b}^{2}}{p_{\rp}\tau} \bI_{N_{\rt}(p+1)}\right]^{-1}\\
\bTheta_{bcu}(p,\alpha) \bydef& [\delta(p,\alpha)\otimes \bR_{bbu}]\bT_{bu}(p,\alpha)[\delta(p,\alpha)\otimes \bR_{bcu}]^{*}\label{eq:Theta}.
\end{align}
Theorem \ref{theorem:channelPrediction} provides the results for the optimal $p$-th order linear Wiener predictor.

\begin{Theorem}
\label{theorem:channelPrediction}
The optimal $p$-th linear Wiener predictor is 
\begin{align}
\bV_{bbu}
=& \alpha [\delta(p,\alpha)\otimes \bR_{bbu}] \bT_{bu}(p,\alpha).\label{eq:V}
\end{align}
\end{Theorem}
\proof
Based on the orthogonality principle~\cite{Haykin1991:BOOK}, $\bV_{bbu}$ can be found by solving the following problem
\begin{align}
\bbE[(\bh_{bbu}[n+1] - \bV_{bbu}\bar{\by}_{\rp,bu}[n])\bar{\by}_{\rp,bu}^{*}[n]]
=& \b0.
\end{align}
Equivalently,
\begin{align} \bbE[\bh_{bbu}[n+1]\bar{\by}^{*}_{\rp,bu}[n]] =& \bV_{bbu}\bbE[\bar{\by}_{\rp,bu}[n]\bar{\by}^{*}_{\rp,bu}[n]].\label{eq:2Dorthogonality}
\end{align}
Thus, the optimal $p$-th order linear Wiener predictor is
\begin{align}
\bV_{bbu}
=& \bR_{\bh\tilde{\bY}}[1]\bR^{-1}_{\tilde{\bY}}[0].\label{eq:V2}
\end{align}
According to~\eqref{eq:Ypt3} and the independence between $\bh_{bbu}$ and $\bh_{bcu}$ for $c\neq b$, the cross-correlation between the true channel and the training signals is
\begin{align}
\bR_{\bh\tilde{\by}}[k+1] 
\bydef&\bbE[\bh_{bbu}[n+1]\tilde{\by}^{*}_{bu}[n-k]]\nn\\
=& \alpha^{|k+1|}\bR_{bbu}.
\end{align}
It follows that
\begin{align}
\bR_{\bh\tilde{\bY}}[1] 
\bydef & \bbE[\bh_{bbu}[n+1]\bar{\by}^{*}_{\rp,bu}[n]]\nn\\
=& \alpha [\delta(p,\alpha) \otimes \bR_{bbu}].\label{eq:RhY}
\end{align}
Moreover, the autocorrelation function of training signals is
\begin{align}
\bR_{\tilde{\by}}[k-q]
\bydef& \bbE[\tilde{\by}_{bu}[n-q]\tilde{\by}^{*}_{bu}[n-k]]\\
=& \alpha^{|k-q|}\sum_{c=1}^{C}\bR_{bcu} + \delta[k-q]\frac{\sigma_{b}^{2}}{p_{\rp}\tau}.\label{eq:Ry}
\end{align}
Consequently,
\begin{align}
\bR_{\tilde{\bY}}[0]
\bydef & \bbE[\bar{\by}_{\rp,bu}[n]\bar{\by}^{*}_{\rp,bu}[n]]\\
=& \begin{pmatrix}
\bR_{\tilde{\by}}[0] 	& 	\bR_{\tilde{\by}}[1] 	& \cdots 	& \bR_{\tilde{\by}}[p]\\
\bR_{\tilde{\by}}[1] 	& 	\bR_{\tilde{\by}}[0] 	& \cdots 	& \bR_{\tilde{\by}}[p-1]\\
\vdots 			& 	 \vdots			& \ddots	& \vdots\\
\bR_{\tilde{\by}}[p]&	\bR_{\tilde{\by}}[p-1] 	& \cdots	& \bR_{\tilde{\by}}[0]\\	
\end{pmatrix}\\
=& \bT^{-1}_{bu}(p,\alpha).\label{eq:RY}
\end{align}
Substituting  \eqref{eq:RhY} and \eqref{eq:RY} into \eqref{eq:V2}, we obtain \eqref{eq:V}. 
\endproof
The predicted channel is 
\begin{align}
\barvh_{bbu}[n+1] 
=& \sum_{q=0}^{p} \bV_{bbu,q}\tilde{\by}_{\rp,bu}[n-q] = \bV_{bbu}\bar{\by}_{\rp,bu}[n].\label{eq:predictedChannel}
\end{align}
The resulting minimum mean squared-error (MMSE) is
\begin{align}
\epsilon_{p}
=& \bbE[||\bh_{bbu}[n+1] - \bV_{bbu}\bar{\by}_{\rp,bu}[n]||_{F}^{2}]\\
=& \tr(\bbE[(\bh_{bbu}[n+1] - \bV_{bbu}\bar{\by}_{\rp,bu}[n])\bh^{*}_{bbu}[n+1]])\\
=& \tr(\bR_{bbu} - \alpha^{2}\bTheta_{bcu}(p,\alpha)).\label{eq:MMSE}
\end{align}
The covariance matrix of $\barvh_{bbu}[n+1] $ is $\alpha^{2}\bTheta_{bbu}(p, \alpha)$. We have the following orthogonal decomposition
\begin{align}
\bh_{bbu}[n+1] &= \barvh_{bbu}[n+1] + \brevevh_{bbu}[n+1],\label{eq:predictedChannelModel}
\end{align}
where $\brevevh_{bbu}[n+1]$ is uncorrelated channel prediction error vector with covariance matrix of $\bR_{bbu} - \alpha^{2}\bTheta_{bcu}(p,\alpha)$. Moreover, we have $\bT_{bu}(0,\alpha) = \bQ_{bu}$ and $\bTheta_{bbu}(0,\alpha) = \bPhi_{bbu}$, thus $\barvh_{bbu}[n+1] = \alpha \hatvh_{bbu}[n]$ when $p = 0$.

\vspace{10pt}
\section{\uppercase{Performance Analysis}}
\label{sec:analysis}
In this section we consider three different scenarios: i) current CSI, ii) aged CSI, and iii) predicted CSI. The superscripts $(\cdot)^{(\rc)}$, $(\cdot)^{(\ra)}$, and $(\cdot)^{(\rp)}$ are used to denote \underline{c}urrent CSI, \underline{a}ged CSI, and \underline{p}redicted CSI.  For both the uplink and the downlink, we first derive achievable SINR expressions for different scenarios for the general setting. Next, we provide some asymptotic results based on the approach using deterministic equivalents in \cite{HoydisEtAl2013:JSAC} for the cases when $N_{\rt}$ is large for MRC receivers on the uplink or MF precoders on the downlink. While these results are asymptotic in the sense that they are derived under an assumption that $N_{\rt} \rightarrow \infty$, simulations in \cite{HoydisEtAl2013:JSAC} show that the fit between simulation and approximation is good, even for small numbers of antennas (around $50$). The approximations are derived under some technical assumptions, that essentially we summarize as (i) the maximum eigenvalue of any spatial correlation matrix is finite, (ii) all spatial correlation matrices have non-zero energy, and (iii) the intercell interference matrix including channel estimation errors are finite. From a practical perspective, the assumptions are reasonable. From the perspective of doing the calculations, only (ii) is problematic. Essentially, one has to remember not to use zero-valued correlation matrices in the expressions. The key ingredient in this asymptotic analysis is the deterministic equivalent SINRs. The resulting expressions are a function of channel covariance matrices $\bR_{b c u}$, the SNR, and various quantities computed from them. We want to emphasize that the analysis does not include overhead, for training or other purposes. Also, we use the notation without temporal index to refer to the deterministic equivalents as $N_{\rt} \rightarrow \infty$. 

\subsection{Uplink Transmission}
\label{subsec:ULanalysis}

Recall that we assume base station $b$ knows $\bR_{bbu}$, $\barvR_{bu}$ for $u \in \cU_{b}$ and $\alpha$. Moreover, depending on the CSI assumption, base station $b$ has the following CSI
\begin{align}
\bg_{bbu}[n+1] 
=& \begin{cases}
\hat{\bh}_{bbu}[n+1], & \mbox{current~CSI}\\
\alpha\hatvh_{bbu}[n], & \mbox{aged~CSI}\\
\barvh_{bbu}[n+1], & \mbox{predicted~CSI}.
\end{cases}
\end{align}
We can rewrite $\tilde{y}_{\rr, bu}[n+1]$ as
\begin{align}
&\tilde{y}_{\rr, bu}[n + 1] 
= \bw_{bu}^{*}[n+1]\bg_{bbu}[n+1]x_{\rr,bu}[n+1]\nn\\
&~~~~~~ + \bw_{bu}^{*}[n+1](\bh_{bbu}[n+1] - \bg_{bbu}[n+1])x_{\rr,bu}[n+1]\nn\\
&~~~~~~ + \sum_{(c,k) \neq (b,u)} \bw_{bu}^{*}[n+1]\bg_{bck}[n+1]x_{\rr,ck}[n+1]\nn\\
&~~~~~~ + \frac{1}{\sqrt{p_{\rr}}}\tilde{z}_{\rr,bu}[n+1].
\end{align} 
Applying the method commonly used in prior work~\cite{Marzetta2006:Asilomar,JoseEtAl2011:TWC,HoydisEtAl2011:Allerton,HoydisEtAl2013:JSAC}, we derive a standard bound on the ergodic achievable uplink rates based on the worst-case uncorrelated additive noise. The idea is to treat $\tilde{y}_{\rr, bu}[n + 1]$ as the  received signal of a single-input single-output (SISO) system with the effective channel of $\bg_{bbu}[n+1]$ while the remaining terms act like uncorrelated additive Gaussian noise. As a result, the desired signal power is 
\begin{align}
S_{\rr, bu} 
=& |\bw_{bu}^{*}[n+1]\bg_{bbu}[n+1]|^{2}.\label{eq:ULgenSignalPower}
\end{align}
The interference plus noise power is
\begin{align}
I_{\rr,bu} 
&= |\bw_{bu}^{*}[n+1](\bh_{bbu}[n+1] - \bg_{bbu}[n+1])|^{2}\nn\\
&+ \frac{\sigma_{b}^{2}}{p_{\rr}}|\bw_{bu}^{*}[n+1]|^{2} + \sum_{(c,k) \neq (b,u)} |\bw_{bu}^{*}[n+1]\bh_{bck}[n+1]|^{2}.\label{eq:ULgenIntfPower}
\end{align}
The post-processed SINR in this case is given by
\begin{align}
\eta_{\rr, bu} 
=& \frac{S_{\rr, bu}}{I_{\rr, bu}}.\label{eq:ULeta}
\end{align}
 The uplink ergodic achievable rate of user $u$ in cell $b$ is
\begin{align}
R_{\rr,bu} 
=& \bbE\left[\log_{2}(1 + \eta_{\rr,bu})\right].\label{eq:ULrate}
\end{align}
Note that the expectation in the expression of the ergodic achievable uplink rate of user $u$ in cell $b$ is over the realizations of the desired channel as in \eqref{eq:ULeta} $\eta_{\rr,bu}$ is computed  for only one realization of the desired channel.

We now consider the asymptotic results when $N_{\rt} \rightarrow \infty$. Lemma \ref{lemma:asymptoticLimits} summarizes the key results used for deriving asymptotic deterministic equivalents in the paper.  Theorem \ref{theorem:ULagedCSI} and Theorem \ref{theorem:ULpredictedCSI} present the expressions of the deterministic equivalent SINR for the aged CSI and predicted CSI for the MRC receiver $\bw_{bu}[n+1] = \bg_{bbu}[n+1]$.

\begin{Lemma}\label{lemma:asymptoticLimits}
Consider $\bA \in \bbC^{N \times N}$ with uniformly bounded spectral norm (with respect to $N$). Consider $\bx$ and $\by$, where $\bx, \by \in \bbC^{N}$, $\bx \sim \cC\cN(\b0, \bPhi_{x})$ and $\by  \sim \cC\cN(\b0, \bPhi_{y})$, are mutually independent and independent of $\bA$. Then, we have
\begin{align}
\frac{1}{N}\bx^{*}\bA\bx - \frac{1}{N}\tr \bA\bPhi_{x} & \xrightarrow[ N \rightarrow \infty]{\mbox{a.s.}} 0 \label{eq:oneVector}\\
\frac{1}{N}\bx^{*}\bA\by & \xrightarrow[ N \rightarrow \infty]{\mbox{a.s.}} 0 \label{eq:twoVector}\\
\bbE\left[\left|\left(\frac{1}{N}\bx^{*}\bA\bx\right)^{2} - \left(\frac{1}{N}\tr \bA \bPhi_{x} \right)^{2} \right|\right] & \xrightarrow[ N \rightarrow \infty]{\mbox{a.s.}}  0\label{eq:squared}\\
\frac{1}{N^{2}} (\bx^{*}\bA\by)^{2} - \frac{1}{N^{2}} \tr \bA \bPhi_{x} \bA \bPhi_{y} &  \xrightarrow[ N\rightarrow \infty]{\mbox{a.s.}} 0. \label{eq:twoVectorGeneral}
\end{align}
\end{Lemma}
\proof
If we denote $\tilde{\bx} = \frac{1}{\sqrt{N}}\bPhi_{x}^{-1/2}\bx$ then $\tilde{\bx} \sim (\b0, \frac{1}{N}\bI_{N})$. Similarly, if we denote $\tilde{\by} = \frac{1}{\sqrt{N}}\bPhi_{y}^{-1/2}\by$ then $\tilde{\by} \sim (\b0, \frac{1}{N}\bI_{N})$. Note that $\tilde{\bx}$ and $\tilde{\by}$ are mutually independent and independent of $\bA, \bPhi_{\rx}$, and $\bPhi_{\ry}$. We can rewrite $\frac{1}{N}\bx^{*}\bA\bx = \tilde{\bx}^{*} \bPhi_{x}^{-1/2}\bA\bPhi_{x}^{-1/2}\tilde{\bx}^{*}$. Applying Lemma 4 (i) in \cite{HoydisEtAl2013:JSAC} for $\tilde{\bx}$ and $\tilde{\bA} = \bPhi_{x}^{1/2}\bA\bPhi_{x}^{1/2}$, we obtain
\begin{align}
\tilde{\bx}^{*} \bPhi_{x}^{1/2}\bA\bPhi_{x}^{1/2}\tilde{\bx} - \frac{1}{N}\tr \bPhi_{x}^{1/2}\bA\bPhi_{x}^{1/2} \xrightarrow[ N \rightarrow \infty]{\mbox{a.s.}} 0.
\end{align}
It follows that
\begin{align}
\frac{1}{N}\bx^{*}\bA\bx - \frac{1}{N}\tr \bA\bPhi_{x} & \xrightarrow[ N \rightarrow \infty]{\mbox{a.s.}} 0,
\end{align}
which is exactly \eqref{eq:oneVector}. Using the same technique, we can prove \eqref{eq:twoVector}, \eqref{eq:squared}, and \eqref{eq:twoVectorGeneral}.
\endproof

\begin{Theorem}\label{theorem:ULagedCSI}
With aged CSI, the deterministic equivalent SINR for user $u$ in cell $b$ is
\begin{align}
\bar{\eta}^{(\ra)}_{\rr,bu}(\alpha) = & \frac{\alpha^{2}A^{(\ra)}_{\rr,bu}}{B^{(\ra)}_{\rr,bu} + C^{(\ra)}_{\rr,bu} + D^{(\ra)}_{\rr,bu} + \alpha^{2}E^{(\ra)}_{\rr,bu}},\label{eq:ULagedCSIeta}
\end{align}
where
\begin{align}
A^{(\ra)}_{\rr,bu} 
=& \left| \tr \bPhi_{bbu} \right|^2\label{eq:AMRC}\\
B^{(\ra)}_{\rr,bu} 
=& \tr (\bR_{bbu} - \alpha^{2}\bPhi_{bbu})\bPhi_{bbu}\label{eq:BMRC}\\
C^{(\ra)}_{\rr,bu} 
=& \frac{\sigma_{b}^{2}}{p_{\rr} } \tr \bPhi_{bbu}\label{eq:CMRC} \\
D^{(\ra)}_{\rr,bu} 
=&  \sum_{(c,k)\neq (,bu)}  \tr \bR_{bck} \bPhi_{bbu}\label{eq:DMRC} \\
E^{(\ra)}_{\rr,bu}
=& \sum_{c \neq b} \left| \tr \bPhi_{bcu} \right|^2.\label{eq:EMRC}
\end{align}
\end{Theorem}
\proof
Substituting $\bg_{bbu}[n+1] = \alpha\hatvh_{bbu}[n]$ into \eqref{eq:ULgenSignalPower}, we obtain the signal power (scaled by $\frac{1}{\alpha^{2}N_{\rt}^{2}}$) as
\begin{align}
S^{(\ra)}_{\rr, bu} 
=& \frac{1}{N_{\rt}^{2}}\alpha^{2} |\hatvh^{*}_{bbu}[n]\hatvh_{bbu}[n]|^{2}. 
\end{align}
Applying Lemma 1, we have
\begin{align}
\frac{1}{N_{\rt}^{2}} |\hatvh^{*}_{bbu}[n]\hatvh_{bbu}[n]|^{2} - \frac{1}{N_{\rt}^{2}}  |\tr \bPhi_{bbu}|^{2} 
\xrightarrow[ N_{\rt} \rightarrow \infty]{\mbox{a.s.}} 0.
\end{align}
Let $\bar{S}^{(\ra)}_{\rr, bu}$ be the deterministic equivalent signal power. Thus, 
\begin{align}
\bar{S}^{(\ra)}_{\rr, bu} - \frac{\alpha^{2}}{N_{\rt}^{2}}  |\tr \bPhi_{bbu}|^{2} \xrightarrow[ N_{\rt} \rightarrow \infty]{\mbox{a.s.}} 0.\label{eq:Theorem2.S}
\end{align}
Substituting $\bg_{bbu}[n+1] = \alpha\hatvh_{bbu}[n]$ and $\bh_{bbu}[n+1] - \bg_{bbu}[n+1] = \tilde{\bee}_{bbu}[n+1]$ into \eqref{eq:ULgenIntfPower}, we obtain the interference plus noise power (scaled by $\frac{1}{\alpha^{2}N_{\rt}^{2}}$) as
\begin{align}
I^{(\ra)}_{\rr,bu}
=& \frac{1}{N_{\rt}^{2}}|\hatvh^{*}_{bbu}[n] \tilde{\bee}_{bbu}[n+1]|^{2} +  \frac{1}{N_{\rt}^{2}}\frac{\sigma_{b}^{2}}{p_{\rr}} \hatvh^{*}_{bbu}[n]\hatvh_{bbu}[n]\nn\\
& + \sum_{(c,k) \neq (b,u)} \frac{1}{N_{\rt}^{2}}\underbrace{|\hatvh^{*}_{bbu}[n]\bh_{bck}[n+1]|^{2}}_{A(c, k , b, u)}.\label{eq:theorem2.I}
\end{align}
Applying Lemma \ref{lemma:asymptoticLimits}, we have
\begin{align}
&\frac{1}{N_{\rt}^{2}} |\hatvh^{*}_{bbu}[n] \tilde{\bee}_{bbu}[n+1]|^{2}\nn\\
& - \frac{1}{N_{\rt}^{2}} \tr \bPhi_{bbu}(\bR_{bbu} - \alpha^{2}\bPhi_{bbu}) \xrightarrow[ N_{\rt} \rightarrow \infty]{\mbox{a.s.}} 0 \label{eq:theorem2.I.1}\\
&\frac{1}{N_{\rt}} \hatvh^{*}_{bbu}[n]\hatvh_{bbu}[n] - \frac{1}{N_{\rt}} \tr \bPhi_{bbu}\xrightarrow[ N_{\rt} \rightarrow \infty]{\mbox{a.s.}} 0.\label{eq:theorem2.I.3}
\end{align}
If $k\neq u$, then $\bh_{bck}[n+1]$ and $\hatvh_{bbu}[n]$ are mutually independent, thus for $k\neq u$
\begin{align}
\frac{1}{N_{\rt}^{2}} A(c, k , b, u) - \frac{1}{N_{\rt}^{2}} \tr \bR_{b c k} \bPhi_{bbu} \xrightarrow[ N_{\rt} \rightarrow \infty]{\mbox{a.s.}} 0.\label{eq:theorem2.I.4}
\end{align}
If $k = u$, define 
\begin{align}
\hat{\bz}_{\rp, bcu}[n] \bydef \sum_{c' \neq c} \bh_{bc'u}[n] + \frac{1}{\sqrt{p_{\rp}\tau}}\tilde{\bz}_{\rp,b}[n].
\end{align} 
As a result, $\hat{\bz}_{\rp, bcu}[n] \sim \cC\cN(\b0, \bQ^{-1}_{bu} - \bR_{bcu})$ and $\hat{\bz}_{\rp, bcu}[n]$ is independent of $\bh_{bcu}[n+1]$. It follows from \eqref{eq:MMSEchannelEstimate} and \eqref{eq:GaussMarkoModel} that
\begin{align}
\hat{\bh}_{bbu}^{*}\bh_{bcu}[n+1] 
&= \alpha \hat{\bh}_{bbu}^{*}\bh_{bcu}[n] + \hat{\bh}_{bbu}^{*}[n]\bee_{bcu}[n+1]\\
&=\alpha \bh_{bcu}^{*}[n]\bQ_{bu}\bR_{bbu}\bh_{bcu}[n]\nn\\
&~+ \alpha\hat{\bz}_{\rp,bcu}^{*}\bQ_{bu}\bR_{bbu}\bh_{bcu}[n]\nn\\
&~+ \hat{\bh}_{bbu}^{*}\bee_{bcu}[n+1].
\end{align}
Thus, we have
\begin{align}
&\frac{1}{N_{\rt}^{2}} A(c, u, b, u) -\Big( \alpha^{2}  \lim_{N_{\rt} \rightarrow \infty} \frac{1}{N_{\rt}^{2}} |\bh_{bcu}^{*}[n]\bQ_{bu}\bR_{bbu}\bh_{bcu}[n]|^{2}\nn\\
&+ \alpha^{2} \frac{1}{N_{\rt}^{2}} |\hat{\bz}_{\rp,bcu}^{*}\bQ_{bu}\bR_{bbu}\bh_{bcu}[n]|^{2}\nn\\
&+ \frac{1}{N_{\rt}^{2}} |\hat{\bh}_{bbu}^{*}\bee_{bcu}[n+1]|^{2}\Big)
\xrightarrow[ N_{\rt} \rightarrow \infty]{\mbox{a.s.}} 0.\label{eq:theorem2.1}
\end{align}
Applying Lemma 1, we have
\begin{align}
& \frac{1}{N_{\rt}^{2}} |\bh_{bcu}^{*}[n]\bQ_{bu}\bR_{bbu}\bh_{bcu}[n]|^{2}  - \frac{1}{N_{\rt}^{2}} |\tr \bPhi_{bcu}|^{2} \xrightarrow[ N_{\rt} \rightarrow \infty]{\mbox{a.s.}} 0\label{eq:theorem2.2}\\
& \frac{1}{N_{\rt}^{2}} |\hat{\bz}_{\rp,bcu}^{*}\bQ_{bu}\bR_{bbu}\bh_{bcu}[n]|^{2}\nn\\
&- \frac{1}{N_{\rt}^{2}} (\tr \bPhi_{bcu}\bR_{bbu} - \tr \bQ_{bu}\bR_{bbu}\bPhi_{bbu}\bR_{bcu}) 
\xrightarrow[ N_{\rt} \rightarrow \infty]{\mbox{a.s.}} 0\label{eq:theorem2.3}\\
&\frac{1}{N_{\rt}^{2}} |\hat{\bh}_{bbu}^{*}\bee_{bcu}[n+1]|^{2} - \frac{(1-\alpha^{2})}{N_{\rt}^{2}} \tr \bPhi_{bcu} \bR_{bbu} \xrightarrow[ N_{\rt} \rightarrow \infty]{\mbox{a.s.}} 0.\label{eq:theorem2.4}
\end{align}
It follows from \eqref{eq:theorem2.1}, \eqref{eq:theorem2.2}, \eqref{eq:theorem2.3}, and \eqref{eq:theorem2.4}, that 
\begin{align}
& \frac{1}{N_{\rt}^{2}} A(c, u, b, u) - \frac{1}{N_{\rt}^{2}} \Big(\alpha^{2}|\tr \bPhi_{bcu}|^{2} + \tr \bR_{bcu}\bPhi_{bbu}\nn\\
& - \alpha^{2} \tr \bQ_{bu}\bR_{bbu}\bPhi_{bbu}\bR_{bcu}\Big)
\xrightarrow[ N_{\rt} \rightarrow \infty]{\mbox{a.s.}} 0.\label{eq:theorem2.I.5}
\end{align}
Let $\bar{I}^{(\ra)}_{\rr, bu}$ be the deterministic equivalent interference plus noise power. From \eqref{eq:theorem2.I}, \eqref{eq:theorem2.I.1}, \eqref{eq:theorem2.I.3},  \eqref{eq:theorem2.I.4}, and \eqref{eq:theorem2.I.5}, we have
\begin{align}
&\bar{I}^{(\ra)}_{\rr, bu} 
- \frac{1}{N_{\rt}^{2}}\Big(\frac{\sigma_{b}^{2}}{p_{\rr}} \tr \bPhi_{bbu} + \alpha^{2} \tr \bPhi_{bbu}(\bR_{bbu} - \bPhi_{bbu})\nn\\
& + (1-\alpha^{2}) \tr \bPhi_{bbu} \bR_{bbu} +  \sum_{(c,k)\neq (b,u)} \tr \bR_{b c k} \bPhi_{bbu}\nn\\
& + \alpha^{2}\sum_{c\neq b} |\tr \bPhi_{bcu}|^{2} - \alpha^{2}\sum_{c\neq b} \tr \bQ_{bu}\bR_{bbu}\bPhi_{bbu}\bR_{bcu} \Big)\nn\\
&\xrightarrow[ N_{\rt} \rightarrow \infty]{\mbox{a.s.}} 0.\label{eq:Theorem2.I.S}
\end{align}
Applying \eqref{eq:Theorem2.I.S} and \eqref{eq:Theorem2.S} into \eqref{eq:ULeta} after neglecting the terms that vanish asymptotically, we obtain $\bar{\eta}^{(\ra)}_{\rr,bu}$ as in \eqref{eq:ULagedCSIeta}.
\endproof

The four terms in the denominator of $\bar{\eta}^{(\rc)}_{\rr,bu}$ characterize the following effects: i) $B^{(\ra)}_{\rr,bu}$ for channel estimation error, ii) $C^{(\ra)}_{\rr,bu}$ for post-processed local noise at base station $b$, iii) $D^{(\ra)}_{\rr,bu}$ for post-processed traditional intra-cell and inter-cell interference, and iv) $\alpha^{2}E^{(\ra)}_{\rr,bu}$ for inter-cell interference due to pilot contamination. Because the current CSI scenario is a special case of the aged CSI scenario when $\alpha = 1$, the deterministic equivalent SINR with current CSI is $\bar{\eta}^{(\rc)}_{\rr,bu} = \bar{\eta}^{(\ra)}_{\rr,bu}(1)$, which is the result in Theorem 3 in ~\cite{HoydisEtAl2013:JSAC}. Moreover, $\bar{\eta}^{(\ra)}_{\rr,bu}(\alpha) \leq \bar{\eta}^{(\rc)}_{\rr,bu}$ since $|\alpha| \leq 1$ and $A^{(\ra)}_{\rr,bu}, B^{(\ra)}_{\rr,bu}, C^{(\ra)}_{\rr,bu}$, and  $D^{(\ra)}_{\rr,bu}$ are nonnegative.  We notice that channel aging affects only desired signal, channel estimation error, and inter-cell interference due to pilot contamination. Specifically, we can show that $\bar{\eta}^{(\ra)}_{\rr,bu}$ is an increasing function of $\alpha$ in $[0,1]$. This means that when user $u$ in cell $b$ moves faster, i.e. $\alpha$ increases, the post-processed uplink SINR is degraded more. Intuitively, when user $u$ in cell $b$ moves, this movement not only affects the desired channel to base station $b$ but also affects the interference channels corresponding to the users sharing the same pilot in other cells, i.e. user $u$ in cells $c \neq b$. Quantitatively, this movement decreases both the desired signal power and the inter-cell interference power due to pilot contamination $\alpha^{2}$ times in the relative comparison with the no channel aging case. Nevertheless, this movement does not affect the local noise power and the \emph{traditional} intra-cell and inter-cell interference from the other users, those do not share the same pilot with user $u$ in cell $b$ during the training stage. This explains how channel aging degrades the uplink ergodic achievable rate.

\begin{Theorem}\label{theorem:ULpredictedCSI}
With predicted CSI obtained by using the optimal $p$-th order linear Wiener predictor, the deterministic equivalent SINR for user $u$ in cell $b$ is
\begin{align}
\bar{\eta}^{(\rp)}_{\rr,bu}(p, \alpha) = & \frac{\alpha^{2}A^{(\rp)}_{\rr,bu}}{B^{(\rp)}_{\rr,bu} + C^{(\rp)}_{\rr,bu} + D^{(\rp)}_{\rr,bu} + \alpha^{2}E^{(\rp)}_{\rr,bu}},\label{eq:ULpredictedCSIeta}
\end{align}
where
\begin{align}
A^{(\rp)}_{\rr,bu} =& |\tr \bTheta_{bbu}(p, \alpha)|^{2}\label{eq:predictedAMRC}\\
B^{(\rp)}_{\rr,bu} =& \tr (\bR_{bbu} - \alpha^{2}\bTheta_{bbu}(p, \alpha))\bTheta_{bbu}(p, \alpha)\label{eq:predictedBMRC}\\
C^{(\rp)}_{\rr,bu} =& \frac{\sigma_{b}^{2}}{p_{\rr}} \tr \bTheta_{bbu}(p, \alpha)\label{eq:predictedCMRC}\\
D^{(\rp)}_{\rr,bu} =& \sum_{(c,k)\neq (b,u)} \tr \bR_{bck}\bTheta_{bbu}(p, \alpha)\label{eq:predictedDMRC}\\
E^{(\rp)}_{\rr,bu} =& \sum_{c\neq b} \alpha^{2p}|\tr \bTheta_{bcu}(p,\alpha)|^{2}.\label{eq:predictedEMRC}
\end{align}
\end{Theorem}
\proof
With predicted CSI, we have $\bg_{bbu}[n+1] = \barvh_{bbu}[n+1]$ and $\bh_{bbu}[n+1] - \barvh_{bbu}[n+1] = \brevevh_{bbu}[n+1]$. Substituting $\bg_{bbu}[n+1] = \barvh_{bbu}[n+1]$ into  \eqref{eq:ULgenSignalPower}, we obtain the signal power (scaled by $\frac{1}{N_{\rt}^{2}}$) as 
\begin{align}
S^{(\rp)}_{\rr, bu} 
=& \frac{1}{N_{\rt}^{2}} |\barvh^{*}_{bbu}[n+1]\barvh_{bbu}[n+1]|^{2}. 
\end{align}
Let $\bar{S}^{(\rp)}_{\rr, bu}$ be the deterministic equivalent signal power. Applying Lemma 1, we have
\begin{align}
&\frac{1}{N_{\rt}^{2}} |\barvh^{*}_{bbu}[n+1]\barvh_{bbu}[n+1]|^{2}\nn\\
& - \frac{1}{N_{\rt}^{2}}  \alpha^{4}|\tr \bTheta_{bbu}(p, \alpha)|^{2} \xrightarrow[ N_{\rt} \rightarrow \infty]{\mbox{a.s.}} 0, 
\end{align}
where $\bTheta_{bbu}(p, \alpha)$ is given in \eqref{eq:Theta}. Thus, 
\begin{align}
\bar{S}^{(\rp)}_{\rr, bu}  - \alpha^{4}|\tr \bTheta_{bbu}(p, \alpha)|^{2} 
\xrightarrow[ N_{\rt} \rightarrow \infty]{\mbox{a.s.}} 0.\label{eq:Theorem3.S}
\end{align}
From \eqref{eq:ULgenIntfPower} and since  $\brevevh_{bbu}[n+1]$ is independent of $\barvh_{bbu}[n+1]$, we obtain the interference plus noise power (scaled by $\frac{1}{N_{\rt}^{2}}$)  as
\begin{align}
I^{(\rp)}_{\rr,bu} 
=& \frac{1}{N_{\rt}^{2}}|\barvh^{*}_{bbu}[n+1] \brevevh_{bbu}[n+1]|^{2} + \frac{1}{N_{\rt}^{2}}\frac{\sigma_{b}^{2}}{p_{\rr}}|\barvh_{bbu}[n+1]|^{2}\nn\\
&+ \sum_{(c,k) \neq (b,u)} \frac{1}{N_{\rt}^{2}}\underbrace{|\barvh^{*}_{bbu}[n+1]\bh_{bck}[n+1]|^{2}}_{B(c,k,b,u)}. \label{eq:Theorem3.1}
\end{align}
Applying Lemma \ref{lemma:asymptoticLimits}, we have
\begin{align}
&\frac{1}{N_{\rt}^{2}}  |\barvh^{*}_{bbu}[n+1] \brevevh_{bbu}[n+1]|^{2}\nn\\
& -\frac{1}{N_{\rt}^{2}} \alpha^{2} \tr  (\bR_{bcu} - \alpha^{2}\bTheta_{bbu}(p, \alpha))\bTheta_{bbu}(p, \alpha)
\xrightarrow[ N_{\rt} \rightarrow \infty]{\mbox{a.s.}} 0\label{eq:Theorem3.2}
\end{align}
\vspace{-15pt}
\begin{align}
\frac{1}{N_{\rt}^{2}}  |\barvh_{bbu}[n+1]|^{2} - \frac{1}{N_{\rt}^{2}} \alpha^{2} \tr \bTheta_{bbu}(p, \alpha)
\xrightarrow[ N_{\rt} \rightarrow \infty]{\mbox{a.s.}} 0.\label{eq:Theorem3.3}
\end{align}
If $k\neq u$, then $\barvh_{bbu}[n+1]$ and $\bh_{bck}[n+1]$ are mutually independent. Thus,
\begin{align}
\frac{1}{N_{\rt}^{2}} B(c,k,b,u) - \frac{1}{N_{\rt}^{2}} \alpha^{2} \tr \bR_{bck} \bTheta_{bbu}(p, \alpha)
\xrightarrow[ N_{\rt} \rightarrow \infty]{\mbox{a.s.}} 0.\label{eq:Theorem3.4}
\end{align}
If $k=u$, it follows from \eqref{eq:predictedChannel} and \eqref{eq:Ypt3} that
\begin{align}
\barvh_{bbu}[n+1]
=& \bV_{bbu}\left(\sum_{c' =1}^{C} \bh_{bc'u}[n] + \frac{1}{\sqrt{p_{\rp}\tau}} \tilde{\bz}_{\rp,b}[n]\right).
\end{align}
As a result, $\barvh_{bbu}[n+1]$ and $\bh_{bcu}[n+1]$ are not mutually independent. We have
\begin{align}
B(c, u, b, u)
=& |\bar{\by}^{*}_{\rp,bu}[n]\bV_{bbu}^{*}\bh_{bcu}[n+1]|^{2}\\
=& \left|\left(\sum_{m=0}^{p}\tilde{\by}^{*}_{bu}[n-m]\bV^{*}_{bbu,m}\right)\bh_{bcu}[n+1]\right|^{2}.\label{eq:Bcubu}
\end{align}
It follows from \eqref{eq:GaussMarkoModel} that
\begin{align}
\bh_{bcu}[n+1] 
=& \alpha^{p+1}\bh_{bcu}[n-p] + \bv_{bcu}[n+1],\label{eq:hbcuDecomposition}
\end{align}
where $\bv_{bcu}[n+1] \sim \cC\cN(\b0,(1 - \alpha^{2(p+1)})\bR_{bcu})$. Also, it follows from \eqref{eq:Ypt3} and \eqref{eq:hbcuDecomposition} that for $m=0, 1, \cdots, p$
\begin{align}
\tilde{\by}^{*}_{bu}[n-m]
=& \alpha^{p-m}\bh^{*}_{bcu}[n-p]+ \sum_{q=0}^{p-m} \alpha^{q}\bee_{bcu}[n-m-q]\nn\\
&+ \hat{\bz}_{\rp,bcu}[n-m].\label{eq:ytildeDecomposition}
\end{align}
Substituting \eqref{eq:hbcuDecomposition} and \eqref{eq:ytildeDecomposition} into \eqref{eq:Bcubu}, then applying Lemma \ref{lemma:asymptoticLimits} and removing asymptotically negligible terms, we obtain
\begin{align}
&\frac{1}{N_{\rt}^{2}} B(c, u, b, u) -  \frac{\alpha^{2(p+1)} }{N_{\rt}^{2}}\nn\\
& \times  \left|\bh^{*}_{bcu}[n-p]\left(\sum_{m=0}^{p}\alpha^{p-m}\bV^{*}_{m}\right)\bh_{bcu}[n-p]\right|^{2}
\xrightarrow[ N_{\rt} \rightarrow \infty]{\mbox{a.s.}} 0.
\end{align}
It follows that
\begin{align}
\frac{1}{N_{\rt}^{2}} B(c, u, b, u) - \frac{\alpha^{2(p+1)}}{N_{\rt}^{2}} |\tr \bTheta_{bcu}(p,\alpha)|^{2}
&\xrightarrow[ N_{\rt} \rightarrow \infty]{\mbox{a.s.}} 0.\label{eq:Theorem3.5}
\end{align}
Applying  \eqref{eq:Theorem3.1}, \eqref{eq:Theorem3.2}, \eqref{eq:Theorem3.3}, \eqref{eq:Theorem3.4},  \eqref{eq:Theorem3.5}, and \eqref{eq:Theorem2.S} into \eqref{eq:ULeta}, we obtain $\bar{\eta}^{(\rp)}_{\rr,bu}$ as in \eqref{eq:ULpredictedCSIeta}.
\endproof
Note that $A^{(\rp)}_{\rr,bu}, B^{(\rp)}_{\rr,bu}, C^{(\rp)}_{\rr,bu}, D^{(\rp)}_{\rr,bu}$, and  $E^{(\rp)}_{\rr,bu}$ in \eqref{eq:ULpredictedCSIeta} have the same role as $A_{\rr,bu}, B_{\rr,bu}, C_{\rr,bu}, D_{\rr,bu}$, and  $E_{\rr,bu}$ in \eqref{eq:ULagedCSIeta}, respectively. Notably, we have $\bar{\eta}^{(\rp)}_{\rr,bu}(0, \alpha) = \bar{\eta}^{(\ra)}_{\rr,bu}(\alpha)$.

\subsection{Downlink Transmission}
\label{subsec:DLanalysis}

We assume that  the users do not have any information on instantaneous channels on the downlink. We use the technique developed in~\cite{Marzetta2006:Asilomar}, which is also used in~\cite{JoseEtAl2011:TWC,HoydisEtAl2013:JSAC},  to derive an expression for downlink achievable rates. Specifically, user $u$ in cell $b$ knows only $\bbE[\bh^{*}_{bbu}[n+1]\bff_{bu}[n+1]]$. Similar to the uplink analysis, we  consider the worst-case uncorrelated additive noise. From \eqref{eq:DLreceivedSignal}, the received signal at user $u$ in cell $b$ (scaled by $1/\sqrt{p_{\rf}}$), is rewritten as
\begin{align}
&y_{\rf,bu}[n+1] =  \sqrt{\lambda_{b}}\bbE[\bh^{*}_{bbu}[n+1]\bff_{bu}[n+1]] x_{\rf,bu}[n + 1]\nn\\
&~~~~~+ \sqrt{\lambda_{b}}(\bh^{*}_{bbu}[n+1]\bff_{bu}[n+1]\nn\\
&~~~~~- \bbE[\bh^{*}_{bbu}[n+1]\bff_{bu}[n+1])x_{\rf,bu}[n + 1]\nn\\
&~~~~~+ \frac{1}{p_{\rf}}z_{\rf,bu}[n+1]\nn\\
&~~~~~+  \sum_{(c,k) \neq (b,u)} \sqrt{\lambda_{c}}\bh^{*}_{cbu}[n+1]\bff_{ck}[n+1]x_{\rf,ck}[n+1].
\end{align}

The signal power at user $u$ in cell $b$ (scaled by $1/(N^{2}_{\rt}p_{\rf})$) is
\begin{align}
S_{\rf,bu}
=& \frac{1}{N_{\rt}^{2}}\lambda_{b}|\bbE[\bh^{*}_{bbu}[n+1]\bff_{bu}[n+1]]|^{2}.\label{eq:DLgenSignalPower}
\end{align}
The interference plus noise power at user $u$ in cell $b$ (scaled by $1/(N^{2}_{\rt}p_{\rf})$) is 
\begin{align}
I_{\rf,bu}
=&  \frac{1}{N_{\rt}^{2}}\lambda_{b}\var[\bh^{*}_{bbu}[n+1]\bff_{bu}[n+1]] + \frac{1}{N_{\rt}^{2}}\frac{\sigma_{bu}^{2}}{p_{\rf}}\nn\\
&+ \sum_{(c,k)\neq (b,u)} \frac{1}{N_{\rt}^{2}}\lambda_{c}\bbE[|\bh^{*}_{cbu}[n+1]\bff_{ck}[n+1]|^{2}]. \label{eq:DLgenIntfPower}
\end{align}
We obtain the achievable SINR and rate of user $u$ in cell $b$ as
\begin{align}
\eta_{\rf,bu} = & S_{\rf,bu} / I_{\rf,bu}\label{eq:DLeta}\\
R_{\rf,bu} =&  \log_{2}(1+ \eta_{\rf,bu}).
\end{align}

For deterministic equivalent analysis as $N_{\rt} \rightarrow \infty$, we focus only on the MF precoders $\bff_{bu}[n+1] = \bg_{bbu}[n+1]$ for all $b \in \cC$ and $u \in \cU_{b}$. Theorem \ref{theorem:DLagedCSI} and Theorem \ref{theorem:DLpredictedCSI} present the downlink deterministic equivalent SINR at user $u$ in cell $b$ in the aged CSI and predicted CSI scenarios.

\begin{Theorem}\label{theorem:DLagedCSI}
With aged CSI, the downlink deterministic equivalent SINR at user $u$ in cell $b$ is
\begin{align}
\bar{\eta}^{(\ra)}_{\rf, bu}(\alpha)
= \frac{\alpha^{4}A^{(\ra)}_{\rf,bu}}{\alpha^{2}B^{(\ra)}_{\rf,bu} + C^{(\ra)}_{\rf,bu} + \alpha^{2}D^{(\ra)}_{\rf,bu} + \alpha^{4}E^{(\ra)}_{\rf,bu}},\label{eq:DLagedCSIeta}
\end{align}
where
\begin{align}
\bar{\lambda}^{(\ra)}_{b} 
=& \left(\alpha \sum_{u'=1}^{U} \tr \bPhi_{bbu'}\right)^{-1}\\
A^{(\ra)}_{\rf,bu} =& \bar{\lambda}^{(\ra)}_{b}|\tr\bPhi_{bbu}|^{2}\label{eq:agedAMF}\\
B^{(\ra)}_{\rf,bu} =& \bar{\lambda}^{(\ra)}_{b} \tr (\bR_{bbu} - \alpha^{2}\bPhi_{bbu})\bPhi_{bbu}\label{eq:agedBMF}\\ 
C^{(\ra)}_{\rf,bu} =& \frac {\sigma^{2}_{bu}} {p_{\rf}}\label{eq:agedCMF}\\
D^{(\ra)}_{\rf,bu} =& \sum_{(c,k) \neq (b,u)} \bar{\lambda}^{(\ra)}_{c}\tr \bR_{cbu}\bPhi_{cck}\label{eq:agedDMF}\\
E^{(\ra)}_{\rf,bu} =& \sum_{c\neq b} \bar{\lambda}^{(\ra)}_{c} |\tr \bPhi_{cbu}|^{2}.\label{eq:agedEMF}
\end{align}
\end{Theorem}
\proof
With aged CSI,  $\bg_{bbu}[n+1] = \alpha \hatvh_{bbu}[n]$, thus the beamforming vector in this case is $\bff^{(\ra)}_{bu}[n+1]= \alpha \hatvh_{bbu}[n]$ for $b \in \cC$ and $u \in \cU_{b}$. It follows from \eqref{eq:lambda} that
\begin{align}
\lambda^{(\ra)}_{b} =& \frac{1}{\alpha^{2} \bbE\left[\sum_{u=1}^{U} \hatvh^{*}_{bbu}[n]\hatvh_{bbu}[n]\right]}.\label{eq:theorem4.1}
\end{align}
Let $\bar{\lambda}^{(\ra)}_{b} \bydef  \lim_{N_{\rt} \rightarrow \infty} \lambda^{(\ra)}_{b}$ be the deterministic equivalent transmit power normalization factor at base station $b\in\cC$. Using \eqref{eq:theorem4.1} and Lemma 1, we obtain 
\begin{align}
\bar{\lambda}^{(\ra)}_{b}
- \alpha^{-2} \left( \sum_{u=1}^{U} \tr \bPhi_{bbu}\right)^{-1} \xrightarrow[ N_{\rt} \rightarrow \infty]{\mbox{a.s.}} 0.\label{eq:theorem4.4}
\end{align}
Substituting \eqref{eq:GaussMarkov2} and $\bff^{(\ra)}_{bu}[n+1]= \alpha \hatvh_{bbu}[n]$ into \eqref{eq:DLgenSignalPower}, we obtain
\begin{align}
S^{(\ra)}_{\rf, bu}
&= \frac{1}{N_{\rt}^{2}}\lambda^{(a)}_{b} |\alpha\bbE[(\alpha\hat{\bh}^{*}_{bbu}[n] + \tilde{\bee}^{*}_{bbu}[n+1])\hat{\bh}_{bbu}[n]]|^{2}\\
&= \frac{1}{N_{\rt}^{2}}\lambda^{(a)}_{b}\alpha^{4} |\bbE[\hat{\bh}^{*}_{bbu}[n]\hat{\bh}_{bbu}[n]]|^{2}.\label{eq:theorem4.6}
\end{align}
Let $\bar{S}^{(\ra)}_{\rf, bu} \bydef \lim_{N_{\rt} \rightarrow \infty} S^{(\ra)}_{\rf, bu}$ be the deterministic equivalent signal power. Using \eqref{eq:theorem4.4} and Lemma 1, we obtain 
\begin{align}
\bar{S}^{(\ra)}_{\rf, bu}
- \frac{1}{N_{\rt}^{2}}\alpha^{4}\bar{\lambda}^{(\ra)}_{b}|\tr \bPhi_{bbu}|^{2} \xrightarrow[ N_{\rt} \rightarrow \infty]{\mbox{a.s.}} 0.\label{eq:theorem4.5}
\end{align}

Substituting $\bff^{(\ra)}_{bu}[n+1]= \alpha \hatvh_{bbu}[n]$ into \eqref{eq:DLgenIntfPower}, we obtain
\begin{align}
I^{(\ra)}_{\rf,bu} =& \frac{\alpha^{2}\lambda^{(\ra)}_{b}}{N_{\rt}^{2}}\var[\bh^{*}_{bbu}[n+1]\hatvh_{bbu}[n]] + \frac{\sigma_{bu}^{2}}{N_{\rt}^{2}p_{\rf}}\nn\\
& + \sum_{(c,k)\neq (b,u)}\frac{\alpha^{2} \lambda^{(\ra)}_{c}}{N_{\rt}^{2}} \underbrace{\bbE[|\bh^{*}_{cbu}[n+1]\hatvh_{cck}[n]|^{2}]}_{C(c,k,b,u)}.\label{eq:theorem4.3}
\end{align}
Using \eqref{eq:GaussMarkov2}, we have
\begin{align}
&\frac{1}{N_{\rt}^{2}}\var[\bh^{*}_{bbu}[n+1]\hatvh_{bbu}[n]] \nn\\
&-  \frac{1}{N_{\rt}^{2}}\bbE[|\tilde{\bee}^{*}_{bbu}[n+1]\hatvh_{bbu}[n]|^{2}]\xrightarrow[ N_{\rt} \rightarrow \infty]{\mbox{a.s.}} 0.\label{eq:theorem4.3B}
\end{align}
Applying Lemma 1 and then using \eqref{eq:theorem4.3B}, we obtain 
\begin{align}
&\frac{1}{N_{\rt}^{2}}\alpha^{2}\lambda^{(\ra)}_{b}\var[\bh^{*}_{bbu}[n+1]\hatvh_{bbu}[n]] \nn\\
&- \alpha^{2}\bar{\lambda}^{(\ra)}_{b} \frac{1}{N_{\rt}^{2}} \tr (\bR_{bbu} - \alpha^{2}\bPhi_{bbu})\bPhi_{bbu} \xrightarrow[ N_{\rt} \rightarrow \infty]{\mbox{a.s.}} 0.\label{eq:theorem4.7}
\end{align}

If $k \neq u$, then $\bh_{cbu}[n+1]$ and $\hatvh_{cck}[n]$ are mutually independent. Thus, when $k \neq u$, we have
\begin{align}
\frac{1}{N_{\rt}^{2}} C(c,k,b,u)  - \frac{1}{N_{\rt}^{2}} \tr \bR_{cbu}\bPhi_{cck}\xrightarrow[ N_{\rt} \rightarrow \infty]{\mbox{a.s.}} 0.\label{eq:theorem4.8}
\end{align}
If $k = u$, then $\bh_{cbu}[n+1]$ and $\hatvh_{ccu}[n]$ are not mutually independent because both depend on $\bh_{cbu}[n]$. We have
\begin{align}
C(c,u,b,u) =& \bbE[|(\alpha\bh^{*}_{cbu}[n] + \bee_{cbu}[n+1])\hatvh_{ccu}[n]|^{2}]\\
=& \alpha^{2}\bbE[|\bh^{*}_{cbu}[n]\bR_{ccu}\bQ_{cu}(\bh_{cbu}[n] + \bz_{\rp, cbu})|^{2}] \nn\\
&+ \bbE[|\bee_{cbu}[n+1]\hatvh_{ccu}[n]|^{2}]\\
=& \alpha^{2}\bbE[|\bh^{*}_{cbu}[n]\bR_{ccu}\bQ_{cu}\bh_{cbu}[n]|^{2}]\nn\\
& + \alpha^{2}\bbE[|\bh^{*}_{cbu}[n]\bR_{ccu}\bQ_{cu}\bz_{\rp, cbu}|^{2}]\nn\\
& + \bbE[|\bee_{cbu}[n+1]\hatvh_{ccu}[n]|^{2}].
\end{align}
Applying Lemma 1, we obtain
\begin{align}
&\frac{1}{N_{\rt}^{2}}C(c,u,b,u) - \Big(\frac{1}{N_{\rt}^{2}} (\alpha^{2}|\tr \bPhi_{ccu}|^{2} +  \tr \bR_{cbu}\bPhi_{ccu}) \nn\\
&+ \frac{\alpha^{2}}{N_{\rt}^{2}}\tr \bR_{cbu}\bPhi_{cbu}\bQ_{cu}\bR_{ccu}\Big)
\xrightarrow[ N_{\rt} \rightarrow \infty]{\mbox{a.s.}} 0.\label{eq:theorem4.9}
\end{align}
Substituting \eqref{eq:theorem4.4}, \eqref{eq:theorem4.7}, \eqref{eq:theorem4.8}, and \eqref{eq:theorem4.9} into \eqref{eq:theorem4.3}, and then substituting the resulting expression and \eqref{eq:theorem4.5} into \eqref{eq:DLeta}, and , we obtain $\bar{\eta}^{(\ra)}_{\rf, bu}(\alpha)$ as in \eqref{eq:DLagedCSIeta}.
\endproof

The terms in \eqref{eq:DLagedCSIeta} characterize the same effects as their counterpart in \eqref{eq:ULagedCSIeta}. Setting $\alpha = 1$, we obtain the deterministic equivalent SINR for the current CSI scenario as $\bar{\eta}^{(\rc)}_{\rf, bu} = \bar{\eta}^{(\ra)}_{\rf, bu}(1)$, which is the result in Theorem 4 in ~\cite{HoydisEtAl2013:JSAC}. We can show that $\bar{\eta}^{(\ra)}_{\rf, bu}(\alpha)$ is an increasing function of $\alpha$ in $[0,1]$. Finally, we also notice that channel aging affects desired signal, channel estimation error, and inter-cell interference due to pilot contamination on the downlink. 

\begin{Theorem}\label{theorem:DLpredictedCSI}
With predicted CSI, the downlink deterministic equivalent SINR at user $u$ in cell $b$ is
\begin{align}
\bar{\eta}^{(\rp)}_{\rf, bu}(p,\alpha)
= \frac{\alpha^{4}A^{(\rp)}_{\rf,bu}}{\alpha^{2}B^{(\rp)}_{\rf,bu} + C^{(\rp)}_{\rf,bu} + \alpha^{2}D^{(\rp)}_{\rf,bu} + \alpha^{4}E^{(\rp)}_{\rf,bu}},\label{eq:DLpredictedCSIeta}
\end{align}
where
\begin{align}
\bar{\lambda}^{(\rp)}_{b} 
=& \left(\alpha^{2}\sum_{u'=1}^{U} \tr \bTheta_{bbu'}(p,\alpha)\right)^{-1}\\
A^{(\rp)}_{\rf,bu} =& \bar{\lambda}^{(\rp)}_{b} |\tr\bTheta_{bbu}(p,\alpha)|^{2}\label{eq:predictedAMF}\\
B^{(\rp)}_{\rf,bu} =& \bar{\lambda}^{(\rp)}_{b} \tr (\bR_{bbu} - \alpha^{2}\bTheta_{bbu}(p,\alpha))\bTheta_{bbu}(p,\alpha)\label{eq:predictedBMF}\\ 
C^{(\rp)}_{\rf,bu} =& \frac {\sigma^{2}_{bu}} {p_{\rf}}\label{eq:predictedCMF}\\
D^{(\rp)}_{\rf,bu} =& \sum_{(c,k)\neq (b,u)} \bar{\lambda}^{(\rp)}_{c}\tr \bR_{cbu}\bTheta_{cck}(p,\alpha)\label{eq:predictedDMF}\\
E^{(\rp)}_{\rf,bu} =&  \alpha^{2p} \sum_{c\neq b} \bar{\lambda}^{(\rp)}_{c}  |\tr \bTheta_{cbu}(p,\alpha)|^{2}.\label{eq:predictedEMF}
\end{align}
\end{Theorem}
\proof
With predicted CSI, we have $\bg_{bbu}[n+1] = \barvh_{bbu}[n+1]$ and hence $\bff^{(\rp)}_{bbu}[n+1] = \barvh_{bbu}[n+1]$  for $b \in \cC$ and $u \in \cU_{b}$. It follows from \eqref{eq:lambda} that
\begin{align}
\lambda^{(\rp)}_{b} =& \frac{1}{\bbE\left[\sum_{u=1}^{U} \barvh^{*}_{bbu}[n+1]\barvh_{bbu}[n+1]\right]}.\label{eq:theorem5.1}
\end{align}
Let $\bar{\lambda}^{(\rp)}_{b} \bydef \lim_{N_{\rt} \rightarrow \infty} \lambda^{(\rp)}_{b} $ be the deterministic equivalent transmit power normalization factor at base station $b\in\cC$. Using \eqref{eq:theorem5.1} and Lemma 1, we obtain
\begin{align}
\bar{\lambda}^{(\rp)}_{b} - \left(\alpha^{2} \sum_{u=1}^{U} \tr \bTheta_{bbu}(p,\alpha)\right)^{-1}
 \xrightarrow[ N_{\rt} \rightarrow \infty]{\mbox{a.s.}} 0.\label{eq:theorem5.0}
\end{align}
Substituting \eqref{eq:GaussMarkov2} and $\bff^{(\rp)}_{bbu}[n+1] = \barvh_{bbu}[n+1]$ into \eqref{eq:DLgenSignalPower}, we obtain
\begin{align}
S^{(\rp)}_{\rf, bu}
&= \frac{1}{N_{\rt}^{2}}\lambda^{(\rp)}_{b} |\bbE[(\barvh^{*}_{bbu}[n+1] + \brevevh^{*}_{bbu}[n+1])\barvh_{bbu}[n+1]]|^{2}\\
&= \frac{1}{N_{\rt}^{2}}\lambda^{(\rp)}_{b}|\bbE[\barvh^{*}_{bbu}[n+1] \barvh_{bbu}[n+1]]|^{2}.\label{eq:theorem5.2}
\end{align}
Let $\bar{S}^{(\ra)}_{\rf, bu} \bydef \lim_{N_{\rt} \rightarrow \infty} S^{(\ra)}_{\rf, bu}$ be the deterministic equivalent signal power. Using \eqref{eq:theorem5.2} and Lemma 1, we obtain
\begin{align}
\bar{S}^{(\rp)}_{\rf, bu}
- \alpha^{4}\frac{1}{N_{\rt}^{2}} \bar{\lambda}^{(p)}_{b} |\tr \bTheta_{bbu}(p,\alpha)|^{2} 
 \xrightarrow[ N_{\rt} \rightarrow \infty]{\mbox{a.s.}} 0.\label{eq:theorem5.S}
\end{align}
Substituting $\bff^{(\rp)}_{bbu}[n+1] = \barvh_{bbu}[n+1]$ into \eqref{eq:DLgenIntfPower}, we obtain
\begin{align}
I^{(\rp)}_{\rf,bu} =& \frac{\lambda^{(\rp)}_{b}}{N_{\rt}^{2}}\var[\bh^{*}_{bbu}[n+1]\barvh_{bbu}[n+1]] + \frac{\sigma_{bu}^{2}}{N_{\rt}^{2}p_{\rf}}\nn\\
& + \sum_{(c,k)\neq (b,u)} \frac{\lambda^{(\rp)}_{c}}{N_{\rt}^{2}}\underbrace{\bbE[|\bh^{*}_{cbu}[n+1]\barvh_{cck}[n+1]|^{2}]}_{D(c,k,b,u)}.\label{eq:theorem5.3}
\end{align}
Using \eqref{eq:predictedChannelModel}, we have
\begin{align}
&\frac{1}{N_{\rt}^{2}}\var[\bh^{*}_{bbu}[n+1]\barvh_{bbu}[n+1]] \nn\\
&- \frac{1}{N_{\rt}^{2}}\bbE[|\breve{\bh}_{bbu}^{*}[n+1]\barvh_{bbu}[n+1]|^{2}] \xrightarrow[ N_{\rt} \rightarrow \infty]{\mbox{a.s.}} 0.\label{eq:theorem5.3B}
\end{align}
Applying Lemma 1 and then using \eqref{eq:theorem5.3B}, we obtain
\begin{align}
&\frac{\lambda^{(\rp)}_{b}}{N_{\rt}^{2}}\var[\bh^{*}_{bbu}[n+1]\barvh_{bbu}[n+1]]\nn\\
&- \frac{\alpha^{2}\bar{\lambda}^{(\ra)}_{b}}{N_{\rt}^{2}}  \tr (\bR_{bbu} - \alpha^{2}\bTheta_{bbu}(p,\alpha))\bTheta_{bbu}(p,\alpha)\xrightarrow[ N_{\rt} \rightarrow \infty]{\mbox{a.s.}} 0.\label{eq:theorem5.5}
\end{align}
If $k \neq u$, then $\bh_{cbu}[n+1]$ and $\barvh_{cck}[n+1]$ are mutually independent. Thus, 
\begin{align}
\frac{1}{N_{\rt}^{2}} D(c,k,b,u) - \alpha^{2}\frac{1}{N_{\rt}^{2}} \tr \bR_{cbu}\bTheta_{cck}(p,\alpha)\xrightarrow[ N_{\rt} \rightarrow \infty]{\mbox{a.s.}} 0.\label{eq:theorem5.6}
\end{align}
If $k = u$, then $\bh_{cbu}[n+1]$ and $\barvh_{ccu}[n+1]$ are not mutually independent because both depend on $\bh_{ccu}[n]$. We have
\begin{align}
D(c,u,b,u) =& |\bar{\by}^{*}_{\rp,cu}[n]\bV^{*}_{ccu}\bh_{cbu}[n+1]|^{2}\\
=& \left|\left(\sum_{m=0}^{p}\tilde{\by}_{cu}[n-m]\bV^{*}_{ccu,m}\right)\bh_{cbu}[n+1]\right|^{2}.\label{eq:Dcubu}
\end{align}
It follows from \eqref{eq:GaussMarkoModel} that
\begin{align}
\bh_{cbu}[n+1] 
=& \alpha^{p+1}\bh_{cbu}[n-p] + \bv_{cbu}[n+1],\label{eq:hcbuDecomposition}
\end{align}
where $\bv_{cbu}[n+1] \sim \cC\cN(\b0,(1 - \alpha^{2(p+1)})\bR_{cbu})$. Also, it follows from \eqref{eq:Ypt3} and \eqref{eq:hcbuDecomposition} that for $m=0, 1, \cdots, p$
\begin{align}
\tilde{\by}^{*}_{cu}[n-m]
=& \alpha^{p-m}\bh^{*}_{cbu}[n-p]+ \sum_{q=0}^{p-m} \alpha^{q}\bee_{cbu}[n-m-q]\nn\\
&+ \hat{\bz}_{\rp,cbu}[n-m].\label{eq:ytildeDecomposition2}
\end{align}
Substituting \eqref{eq:hcbuDecomposition} and \eqref{eq:ytildeDecomposition2} into \eqref{eq:Bcubu}, then applying Lemma \ref{lemma:asymptoticLimits} and removing asymptotically negligible terms, we obtain
\begin{align}
&\frac{1}{N_{\rt}^{2}} D(c, u, b, u) - \frac{\alpha^{2(p+1)} }{N_{\rt}^{2}}\nn\\
&\times \left|\bh^{*}_{cbu}[n-p]\left(\sum_{m=0}^{p}\alpha^{p-m}\bV^{*}_{ccu,m}\right)\bh_{cbu}[n-p]\right|^{2}\xrightarrow[ N_{\rt} \rightarrow \infty]{\mbox{a.s.}} 0.
\end{align}
Applying Lemma \ref{lemma:asymptoticLimits} and then using \eqref{eq:V} and \eqref{eq:Theta}, we have
\begin{align}
\frac{1}{N_{\rt}^{2}} D(c, u, b, u) -  \frac{\alpha^{2(p+1)}}{N_{\rt}^{2}} |\tr \bTheta_{cbu}(p,\alpha)|^{2}\xrightarrow[ N_{\rt} \rightarrow \infty]{\mbox{a.s.}} 0.\label{eq:theorem5.7}
\end{align}
Substituting \eqref{eq:theorem5.0}, \eqref{eq:theorem5.5}, \eqref{eq:theorem5.6}, and \eqref{eq:theorem5.7} into \eqref{eq:theorem5.3}, and then substituting the resulting expression and \eqref{eq:theorem5.S} into \eqref{eq:DLeta}, we obtain $\bar{\eta}^{(\rp)}_{\rf, bu}(p,\alpha)$ as in \eqref{eq:DLpredictedCSIeta}.
\endproof

Note that $\bar{\eta}^{(\rp)}_{\rf, bu}(0,\alpha) = \bar{\eta}^{(\ra)}_{\rf, bu}(\alpha)$.

\vspace{10pt}
\section{\uppercase{Numerical Results}}
\label{sec:numericalResults}

\begin{figure}
\centerline{
\includegraphics[width=0.9\columnwidth]{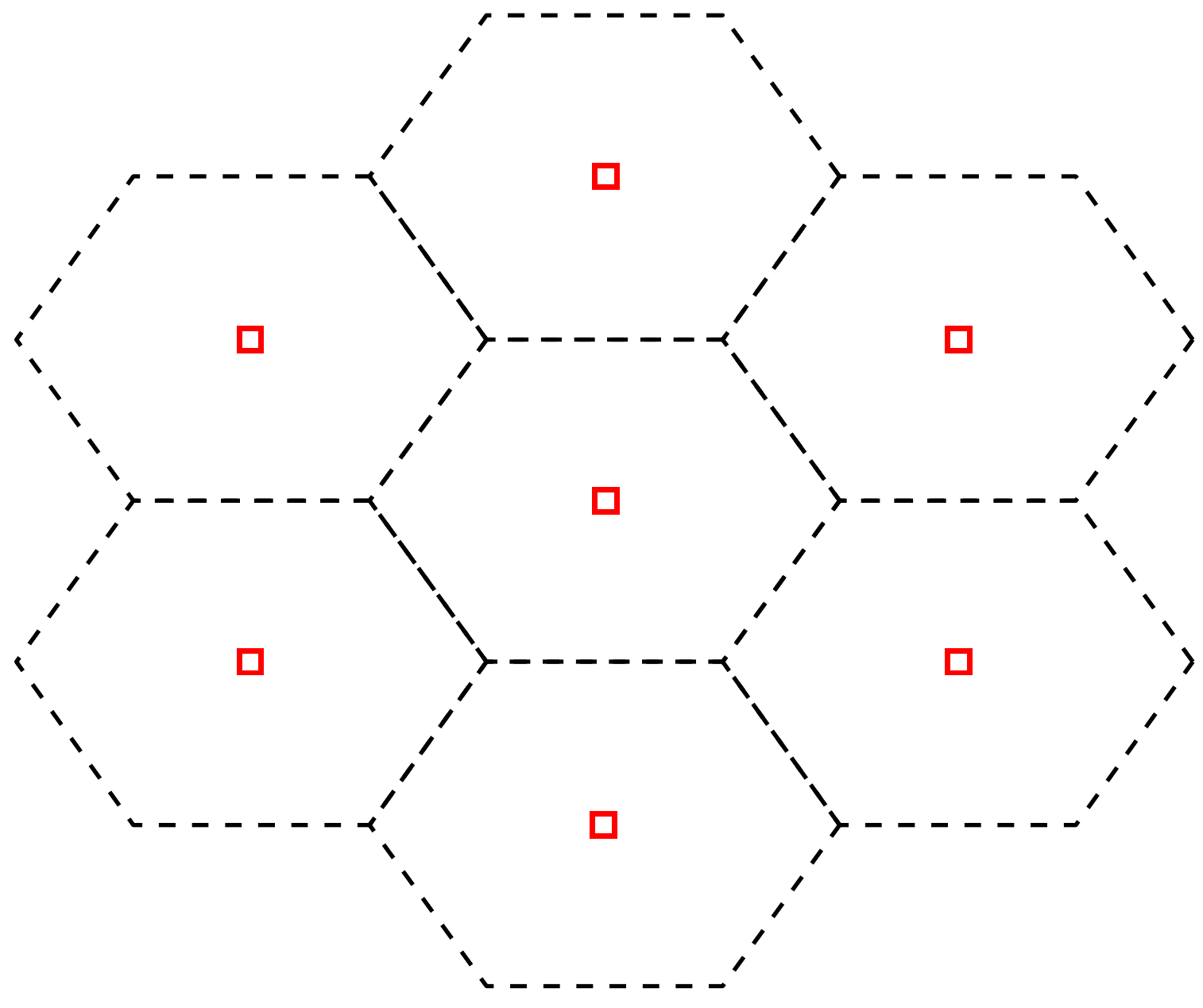}
}
\caption{Base stations are located at the center of cells and are illustrated by the red squares. We are interested in the average achievable sum-rates of the users in the center cell.}
\label{fig:TypicalUserRing}
\end{figure}

In this section, we present numerical results on the effects of channel aging and the benefits of channel prediction in a multi-cell network. Specifically, the simulated network consists of 7 cells as illustrated in \figref{fig:TypicalUserRing}. We assume the cells have the same number of users. Also, we assume that all the cells have the same base station configurations, e.g., the number of antennas at each base station and how the antennas are deployed in the cell.  Several key simulation parameters are provided in Table \ref{tab:simParameters}. This set of parameters correspond to an interference-limited scenario. We are interested in computing the average achievable sum-rates of the users in the center cells on the downlink and on the uplink. Because we notice that similar observations can be made for the uplink and for the downlink,  we present only selected results due to space constraints.

\begin{table}[h]
\caption{Simulation parameters}
\centering
\begin{tabular}{| p{3.5cm} | p{4.5cm} |}
\hline
Parameter & Description \\\hline
Inter-site distance & 500 meters\\\hline
Number of users per cell & 12 \\\hline
Path-loss model & $\mathrm{PLNLOS} =128.1+ 37.6 \log10(d)$ where $d > 0.035$km is the transmission distance in kilometers and $f_c = 2$GHz is the carrier frequency, \\\hline
Penetration loss & 20dB\\\hline
Antenna array configuration at users & 1 antenna omni with 0dBi gain\\\hline
CSI delay & 1ms to 10 ms\\\hline
User dropping & uniformly distributed within a cell\\\hline
Shadowing model & not considered\\\hline
User assignment & each user is served by the base station in the same cell\\\hline
BS antenna gain & 10dBi\\\hline
BS total transmit power & 46dBm\\\hline
UE speed & 3 to 120 kmh\\\hline
Frequency carrier & 2GHz\\\hline
Bandwidth & 10MHz\\\hline
Thermal noise density & -174dBm/Hz\\\hline
BS noise figure & 5dB\\\hline
UE noise figure & 9dB \\\hline 
\end{tabular}\label{tab:simParameters}
\end{table}


To incorporate spatial correlation in the simulations, we consider a circular array geometry for massive MIMO with a single scattering cluster, of fixed spread, and randomly chosen. Moreover, the spatial correlation is chosen to be independent of the temporal correlation so that we can separate the channel aging and spatial correlation effects.  Indeed, there are many potential array geometries for massive MIMO. While patch antennas seem attractive, other miniaturized antenna designs may alternatively be employed (patch antennas are not used on mobile devices for example). At this point, there is no defacto geometry and channel model that is considered a standard for massive MIMO simulations. We chose to use a circular array because they are commercially deployed in other systems that exploit reciprocity, i.e. TD-SCDMA in China, and it is possible to compute the spatial correlation matrix efficiently using algorithms developed in prior work~\cite{ForenzaEtAl2007:TVT}. Specifically, the approach in~\cite{ForenzaEtAl2007:TVT}  provides a closed-form approximate expression of the spatial correlation matrix between a user and a base station. In the following simulations, we consider mainly the uniformly circular array (UCA) configuration and assume that the antenna array configuration has a uniformly chosen angle-of-arrival (or departure), and a given angle spread (AoS) $\sigma_{\ras}$. Moreover, we assume that the angles of arrival (AoAs)/angles of departure (AoDs) are distributed according to a certain power azimuth spectrum (PAS). The PAS is modeled by the truncated Laplacian pdf, which is given by
\begin{align}\label{eq:LaplacianPDF}
P_{\phi}(\phi)
=& \begin{cases}
\frac{\beta_{\ras}}{\sqrt{2}\sigma_{\ras}} e^{-|\sqrt{2} \phi/\sigma_{\ras}|}, & \mbox{~if~} \phi \in [-\pi, \pi]\\
0, & \mbox{otherwise},
\end{cases}
\end{align}
where $\phi$ is the random variable describing the AoA/AoD with respect to the mean angle $\phi_{0}$, and $\beta_{\ras} = \left(1 - e^{-\sqrt{2}\pi / \sigma_{\ras}}\right)^{-1}$. Note that because our simulations also consider random user locations (unlike prior work that usually considers a fixed location of users and interferers), the complexity of computing the spatial correlation matrix becomes the bottleneck in our simulations (according to the MATLAB profile function).  Our approach does work with other more complex antenna and correlation models at the expense of computational complexity.

\begin{figure}
\centerline{
\includegraphics[width=0.9\columnwidth]{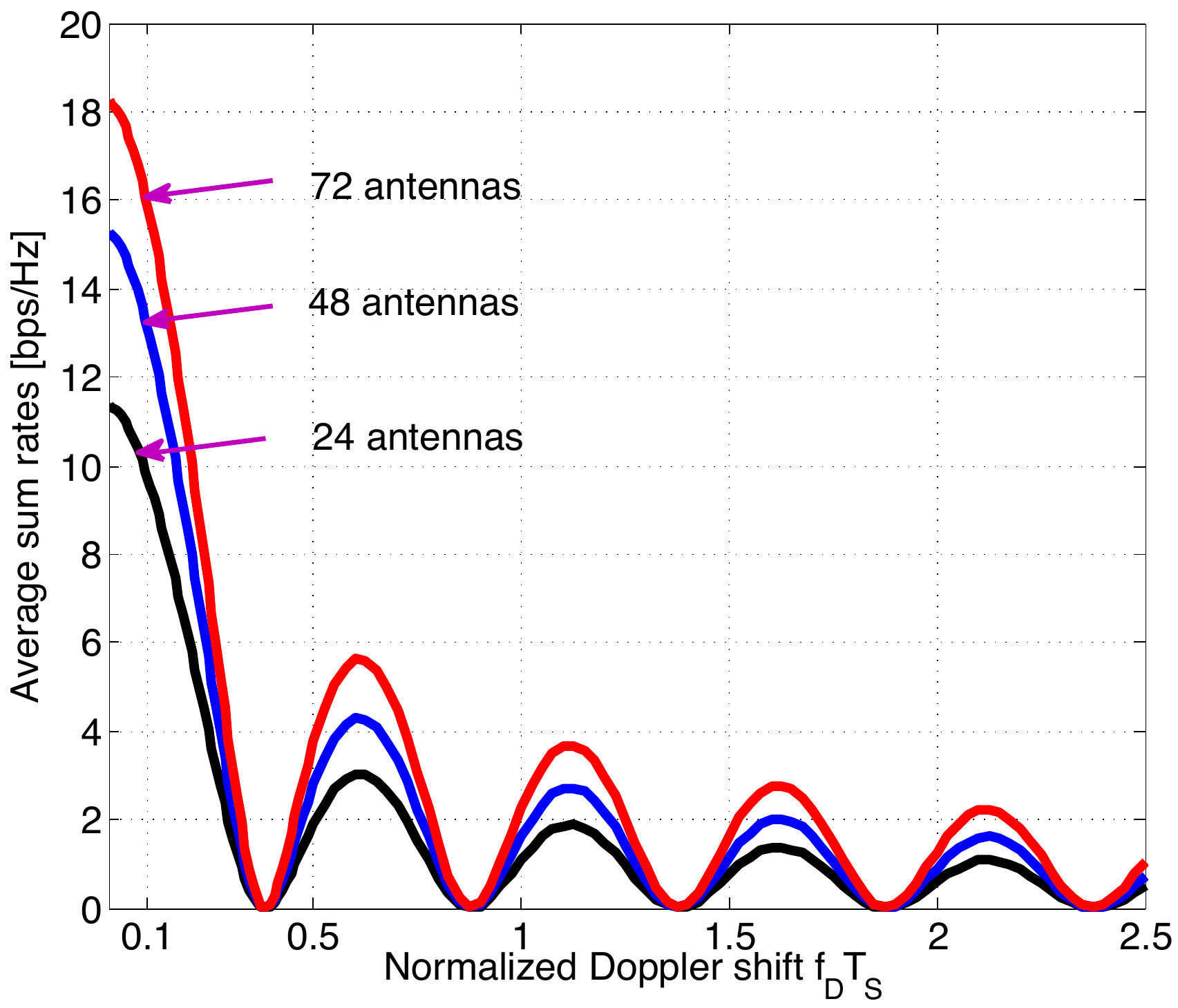}
}
\caption{The downlink average achievable sum-rates of the users in the center cell as a function of normalized Doppler shifts for different numbers of antennas at each base station.}
\label{fig:DLnormalizedDoppler}
\end{figure}

We use normalized Doppler shifts, which are defined as $f_{\rD}T_{s}$, to characterize channel aging. Larger normalized Doppler shifts correspond to large velocities of the users or large CSI delays.  \figref{fig:DLnormalizedDoppler} shows the downlink average achievable sum-rates of the users in the center cell as a function of the normalized Doppler shifts for $N_{\rt} \in \{24, 48, 72\}$. We notice the trend that  the downlink average achievable sum-rates of the users in the center cell decreases in magnitude to zero though not monotonically since there are some ripples. At first it decreases with the increasing $f_{\rD}T_{s}$ until getting to $f_{\rD}T_{s} \approx 0.4$. Moreover, channel aging reduces the downlink average achievable sum-rates of the users in the center cell by half at $f_{\rD}T_{s} \approx 0.2$. Finally, increasing $N_{\rt}$ does not help improve the value of $f_{\rD}T_{s}$ at which  the downlink average achievable sum-rates of the users in the center cell gets to zero for the first time.


\begin{figure}
\centerline{
\includegraphics[width=0.9\columnwidth]{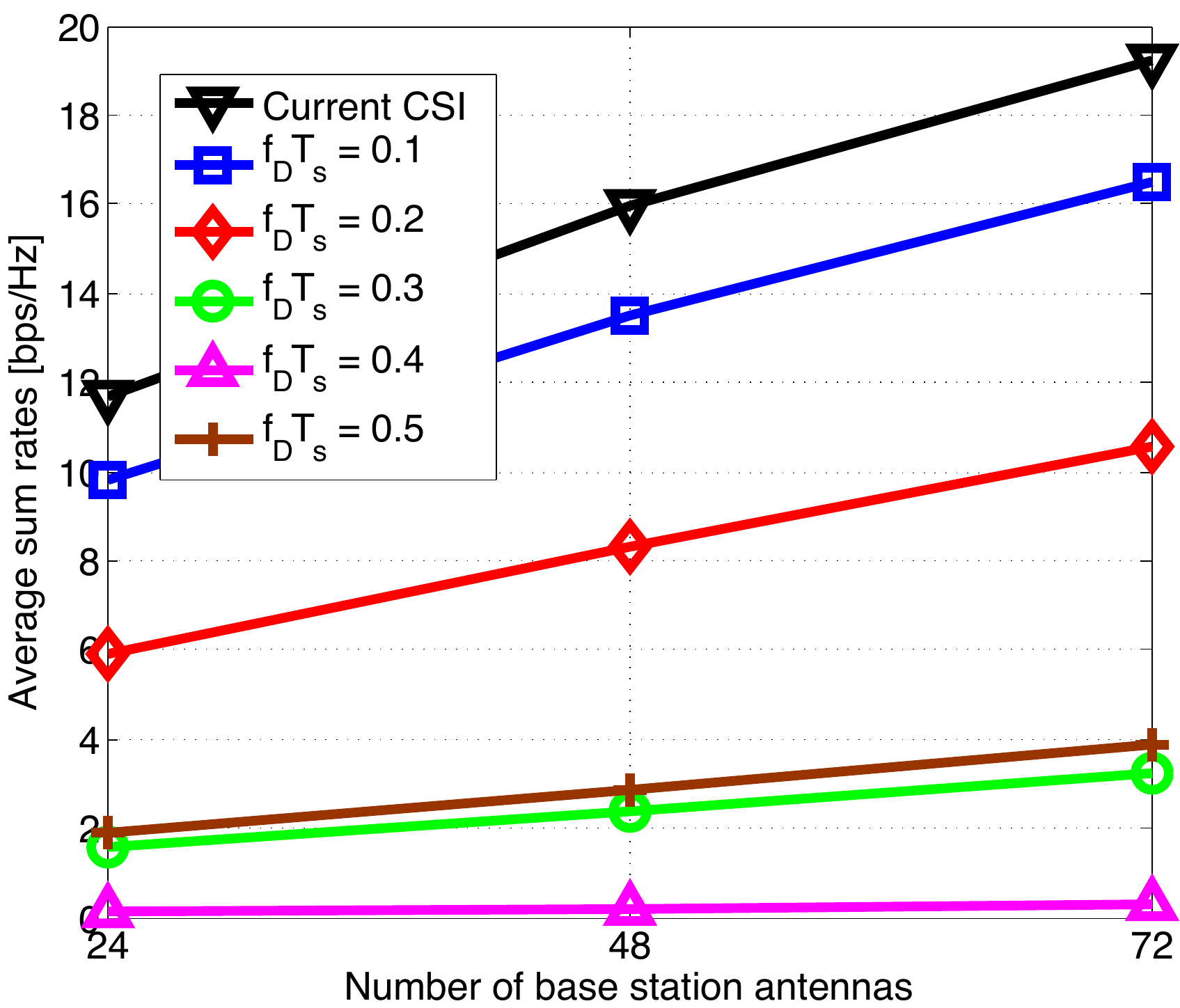}
}
\caption{The uplink average achievable sum-rates of the users in the center cell as a function of the number of antennas at a base station for different normalized Doppler shifts. Each cell has 12 active users, that are uniformly distributed in the cell area.}
\label{fig:ULBSantenna}
\end{figure}

\figref{fig:ULBSantenna} presents the uplink average achievable sum-rates of the users in the center cell as a function of the number of antennas at a base station for different normalized Doppler shifts. We notice that increasing $N_{\rt}$ improves the uplink average achievable sum-rates of the users in the center cell. Moreover, when $f_{\rD}T_{s} = 0.4$, the uplink average achievable sum-rates of the users in the center cell is negligible even for a large number of antennas at a base station, e.g. $N_{\rt} = 72$. Also, we observe that for the simulation setting, the uplink average achievable sum-rates of the users in the center cell when $f_{\rD}T_{s} = 0.2$ is always more than half of that in the case of current CSI.


\begin{figure}
\centerline{
\includegraphics[width=0.9\columnwidth]{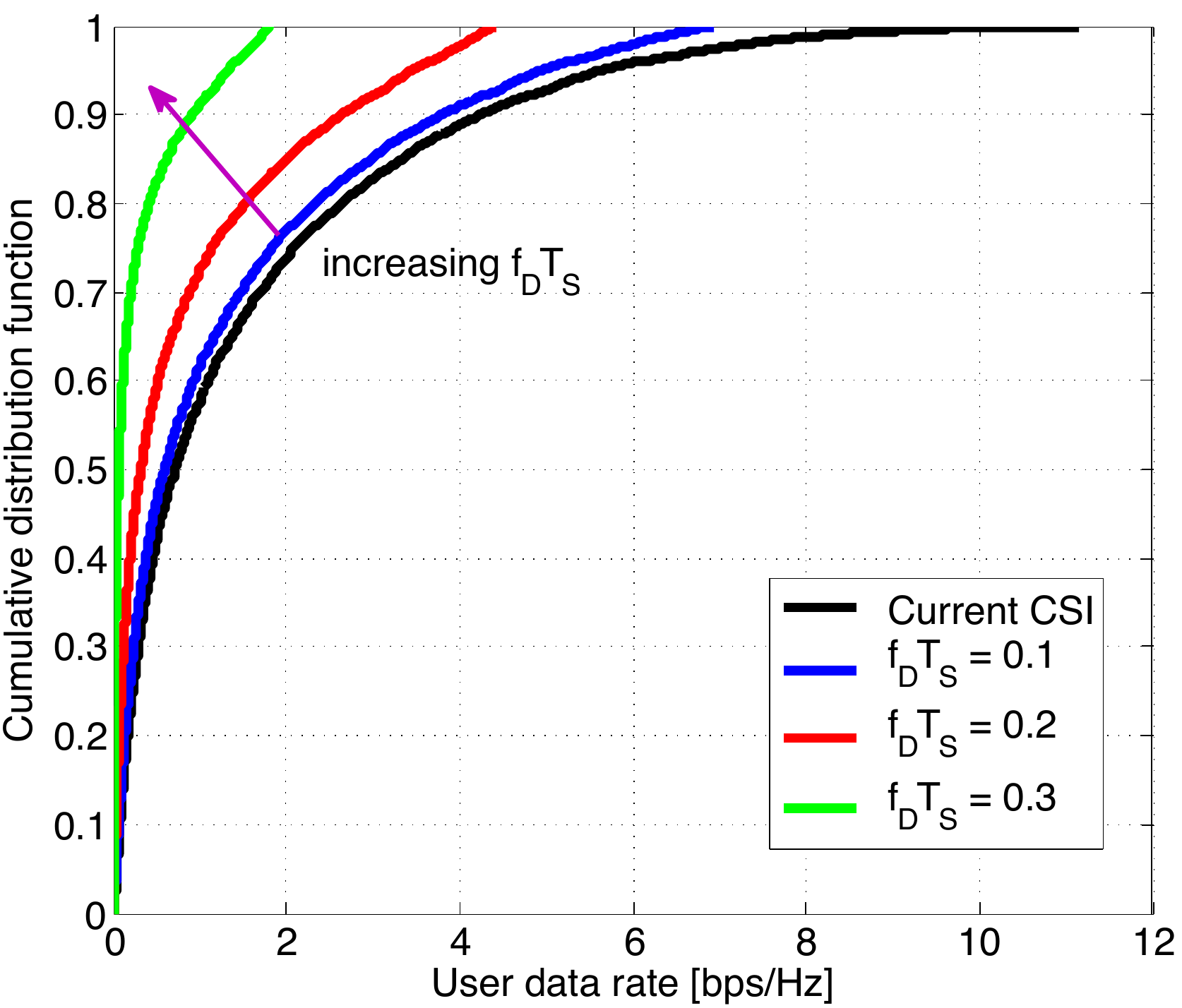}
}
\caption{The cumulative distribution function of the downlink achievable rate of users in the center cell for different normalized Doppler shifts.}
\label{fig:DLrateCDF72antennas}
\end{figure}

In the previous experiments, we consider the achievable sum-rates of the users in the center cell. Fig. \ref{fig:DLrateCDF72antennas} shows the cumulative distribution function (CDF) of the downlink achievable rate of the users in the center cell for different normalized Doppler shifts. Notably, channel aging affects the peak rates significantly. Similar to the other experiments, we notice that the channel aging corresponding to $f_{\rD}T_{s} = 0.2$ gracefully degrades the rate distribution of the users in the center cell.


\begin{figure}
\centerline{
\includegraphics[width=0.95\columnwidth]{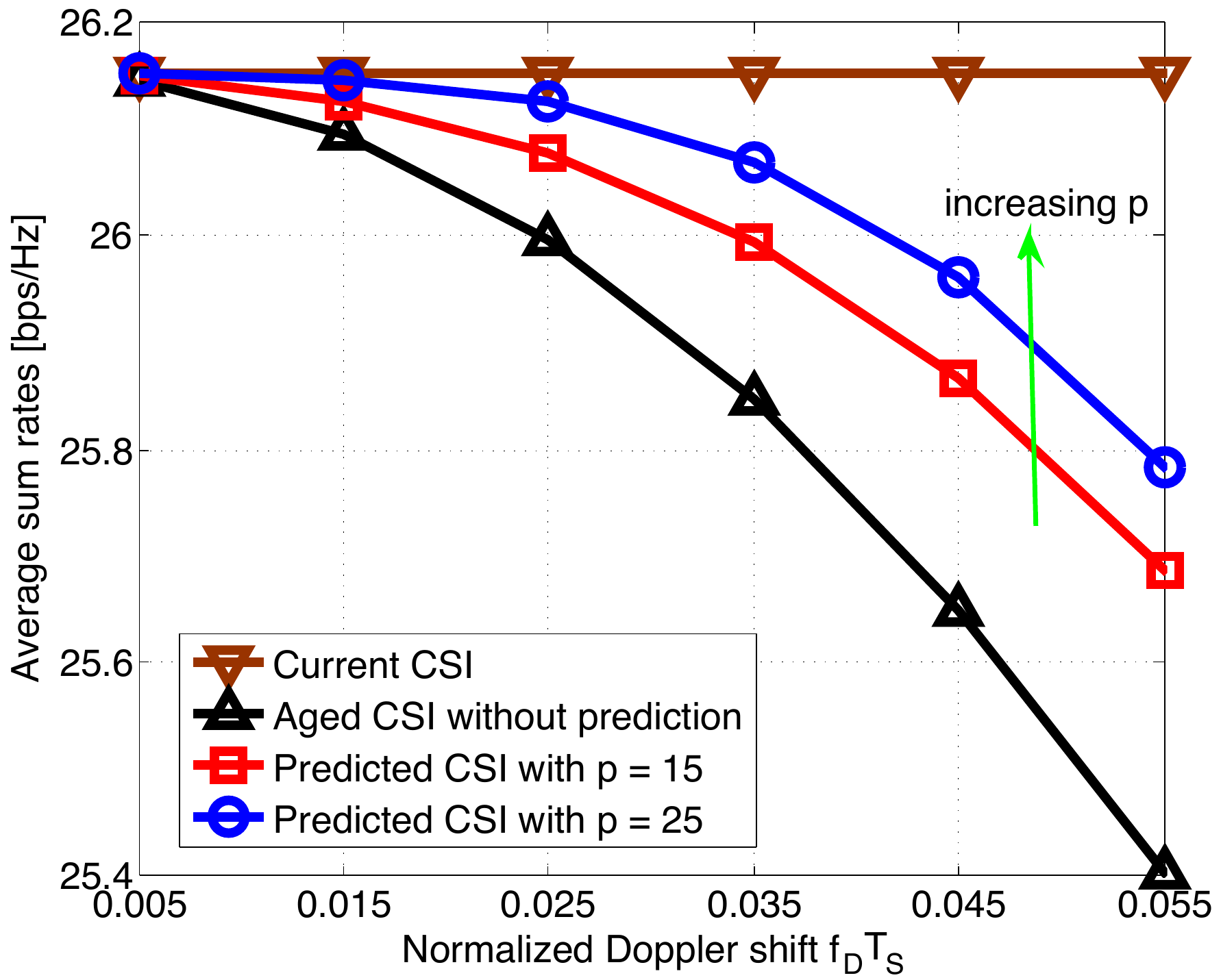}
}
\caption{The uplink achievable rate of the typical user as a function of different normalized Doppler shifts without prediction and with FIR prediction of $p=50$.}
\label{fig:ULchannelPrediction}
\end{figure}

We now investigate the benefits of FIR channel prediction when each base station is equipped with 120 antennas. To reduce the computational complexity associated with spatial correlation matrix computation and to separate between spatial correlation and temporal correlation, in this experiment we consider only the spatially uncorrelated channel model. \figref{fig:ULchannelPrediction} shows the uplink average achievable sum-rates of the users in the center cell as a function of different normalized Doppler shifts without prediction and with FIR prediction of $p=15$ and $p=25$. We notice that channel prediction does help cope with channel aging, however, for these values of $p$, the relative gain is not large, especially at large normalized Doppler shifts. Moreover, the larger value of $p$ makes use of more observations in the past to provide a higher gain. Alternatively, the Kalman filter, which is used to approximate the non-causal Wiener filter, may cope better with channel aging effects~\cite{Hayes1996:BOOK}. The investigation of the Kalman-filter based channel prediction in massive MIMO systems is left for future work.
 

\vspace{10pt}
\section{\uppercase{Conclusion}}
\label{sec:conclusion}

In this paper, we proposed a new framework to incorporating practical effects like channel aging into massive MIMO systems. Channel aging causes the mismatch between the channel when it is learned and that when it is used for beamforming or detection. We derived the optimal causal linear FIR Wiener predictor to overcome the issues.  We also provided analysis of achievable rate performance on the uplink and on the downlink in the presence of channel aging and channel prediction. Numerical results showed that although channel aging degraded performance of massive MIMO, the decay due to aging is graceful. Simulations also showed the potential of channel prediction to overcome channel aging.

The work focuses only on channel aging. Other practical effects in massive MIMO need to be studied as well. For example, the large arrays in massive MIMO are likely to be closely spaced, leading to correlation in the channel. Future work can investigate the impact of spatial correlation on massive MIMO performance, especially the comparison between collocated antennas and distributed antennas. Our initial results along these lines are found in~\cite{TruongHeath2013:Asilomar}. In this paper, we consider only the MRC receivers on the uplink and the MF precoder on the downlink. Thus, another interesting topic is to analyze the performance in when other types of receivers/precoders are used. Finally, future work can investigate the use of more complicated channel predictors to overcome channel aging effects.


\bibliographystyle{jcn}


\epsfysize=3.2cm
\begin{biography}{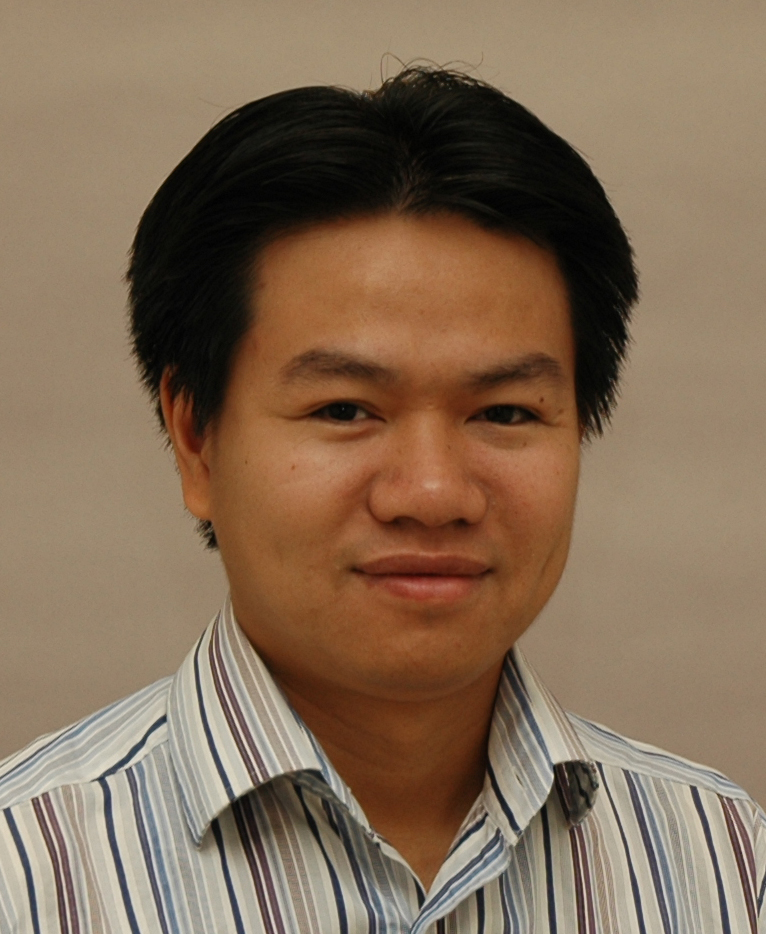}{Kien T. Truong} received the B.S. degree in electronics and telecommunications from Hanoi University of Technology, Hanoi, Vietnam, in 2002, and the M.Sc. and Ph.D. degrees in electrical engineering from The University of Texas at Austin, Austin, TX, USA, in 2008 and 2012, respectively. From 2002 to 2006, he was with the Department of Wireless Communications, Research Institute of Posts and Telecommunications, Hanoi, Vietnam. He was a 2006 Vietnam Education Foundation (VEF) Fellow. He is a Consultant at MIMO Wireless Inc. His research interests include massive MIMO communication, link adaptation and interference management for wireless cooperative communications, and capacity analysis of wireless ad hoc networks. He was co-recipient of the 2013 EURASIP Journal on Wireless Communications and Networking best paper award. 
\end{biography}

\epsfysize=3.2cm
\begin{biography}{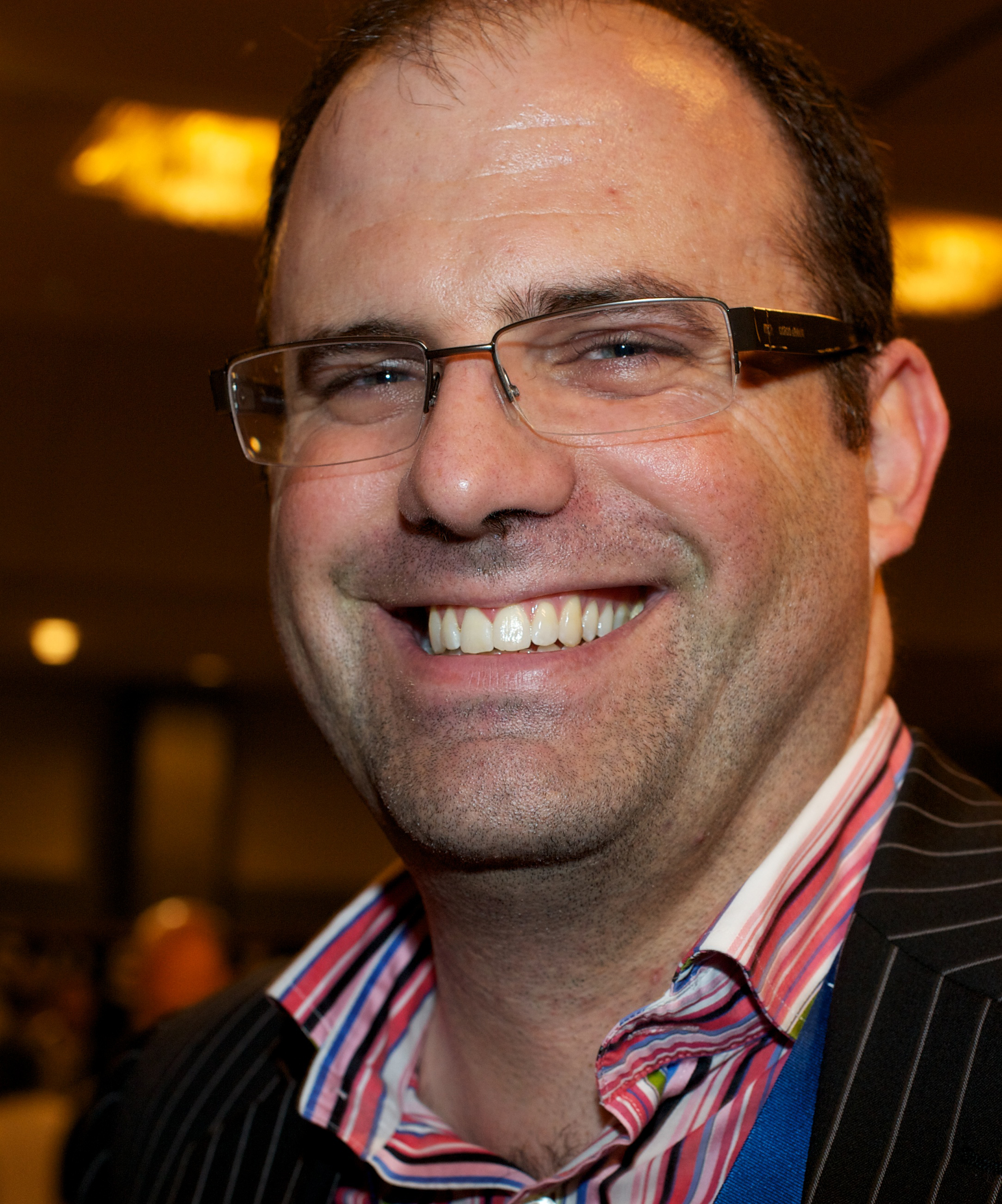}{Robert W. Heath Jr.}  received the B.S. and M.S. degrees
from the University of Virginia, Charlottesville, VA, in 1996 and 1997 respectively, and the Ph.D. from Stanford University, Stanford, CA, in 2002, all in electrical engineering. From 1998 to 2001, he was a Senior Member of the Technical Staff then a Senior Consultant at Iospan Wireless Inc, San Jose, CA where he worked on the design and implementation of the physical and link layers of the first commercial MIMO-OFDM communication system. Since January 2002, he has been with the Department of Electrical and Computer Engineering at The University of Texas at Austin where he is a Professor and Director of the Wireless Networking and Communications Group. He is also President and CEO of MIMO Wireless Inc. and Chief Innovation Officer at Kuma Signals LLC. His research interests include several aspects of wireless communication and signal processing: limited feedback techniques, multihop networking, multiuser and multicell MIMO, interference alignment, adaptive video transmission, manifold signal processing, applications of stochastic geometry, and millimeter wave communication techniques. 

Dr. Heath has been an Editor for the IEEE Transactions on Communication, an Associate Editor for the IEEE Transactions on Vehicular Technology,  lead guest editor for an IEEE Journal on Selected Areas in Communications special issue on limited feedback communication, and lead guest editor for an IEEE Journal on Selected Topics in Signal Processing special issue on Heterogenous Networks. He currently serves on the steering committee for the IEEE Transactions on Wireless Communications. He was a member of the Signal Processing for Communications Technical Committee in the IEEE Signal Processing Society and is a former  Chair of the IEEE COMSOC Communications Technical Theory Committee. He was a technical co-chair for the 2007 Fall Vehicular Technology Conference, general chair of the 2008 Communication Theory Workshop, general co-chair, technical co-chair and co-organizer of the 2009 IEEE Signal Processing for Wireless Communications Workshop, local co-organizer for the 2009 IEEE CAMSAP Conference, technical co-chair for the 2010 IEEE International Symposium on Information Theory,  the technical chair for the 2011 Asilomar Conference on Signals, Systems, and Computers, general chair for the 2013 Asilomar Conference on Signals, Systems, and Computers, general co-chair for the 2013 IEEE GlobalSIP conference, and is technical co-chair for the 2014 IEEE GLOBECOM conference.

Dr. Heath was a co-author of best student paper awards at IEEE  VTC 2006 Spring, WPMC 2006, IEEE GLOBECOM 2006, IEEE VTC 2007 Spring, and IEEE RWS 2009, as well as co-recipient of the Grand Prize in the 2008 WinTech WinCool Demo Contest.  He was co-recipient of the 2010 and 2013 EURASIP Journal on Wireless Communications and Networking best paper awards and the 2012 Signal Processing Magazine best paper award. He was a 2003 Frontiers in Education New Faculty Fellow. He is the recipient of the David and Doris Lybarger Endowed Faculty Fellowship in Engineering, an IEEE Fellow, a licensed Amateur Radio Operator, and is a registered Professional Engineer in Texas.\end{biography}

\end{document}